\PassOptionsToPackage{warn}{textcomp}
\documentclass[10pt,journal,compsoc]{IEEEtran}

\usepackage[english]{babel}
\usepackage{blindtext}
\usepackage{booktabs} 
\usepackage{pifont}
\usepackage{array}
\usepackage{tcolorbox} \newcommand{\ciao}[1]{{\setlength\fboxrule{0pt}\fbox{\tcbox[colframe=black,colback=white,shrink tight,boxrule=0.2pt,extrude by=0.5mm]{\small #1}}}}

\newcommand{\cmark}{\ding{51}}%

\newcommand{\ie}{{\em i.e., }}
\newcommand{\eg}{{\em e.g., }}
\newcommand{\RNum}[1]{\uppercase\expandafter{\romannumeral #1\relax}}
\newcommand{\myverb}{\fontsize{10}{48}\usefont{OT1}{lmtt}{b}{n}\noindent }

\usepackage{adjustbox}
\usepackage{subfigure}
\usepackage{balance}
\usepackage{float}

\usepackage{amsmath}
\usepackage{array}
\usepackage{multirow}
\usepackage{bbding}
\usepackage{dblfloatfix}
\usepackage{color, colortbl}
\usepackage{graphicx}
\usepackage{hyperref}

\usepackage[T1]{fontenc}

\begin{document}
	
\title{Detecting Anomalous Microflows in IoT Volumetric Attacks via Dynamic Monitoring of MUD Activity}

\newcommand*{\affmark}[1][*]{\textsuperscript{#1}}

\author{Ayyoob~Hamza,
	Hassan~Habibi~Gharakheili, 
	Theophilus~A.~Benson, 
	Gustavo~Batista,
	and~Vijay~Sivaraman
	\IEEEcompsocitemizethanks{
		\IEEEcompsocthanksitem A. Hamza, H. Habibi Gharakheili, and V. Sivaraman  are with the School of Electrical Engineering and Telecommunications, University of New South Wales, Sydney, NSW 2052, Australia (e-mails: m.ahamedhamza@unsw.edu.au, h.habibi@unsw.edu.au, vijay@unsw.edu.au).
		\IEEEcompsocthanksitem G. Batista is with the School of Computer Science Engineering, University of New South Wales, Sydney, NSW 2052, Australia  (e-mail: g.batista@unsw.edu.au).				
		\IEEEcompsocthanksitem T. Benson is with the School	of Electrical and Computer Engineering Department at Carnegie Mellon University, USA (e-mail: theophilus@cmu.edu).		
		\IEEEcompsocthanksitem This submission is an extended and improved version of our paper presented at the ACM SOSR 2019 \cite{MUDlearn}.
	}
}

\markboth{}%
{Shell \MakeLowercase{\textit{et al.}}: Bare Demo of IEEEtran.cls for Computer Society Journals}
\maketitle
\begin{abstract}

Smart 	homes, enterprises, and cities equipped with IoT devices are increasingly becoming target of an escalating number of sophisticated new cyber-attacks. Anomaly-based detection methods are promising in finding new attacks, but there are certain practical challenges like false-positive alarms, hard to explain, and difficult to scale cost-effectively.
The IETF recent standard called Manufacturer Usage Description (MUD) seems promising to limit the attack surface on IoT devices by formally specifying their intended network behavior (whitelisting). In this paper, we use SDN to enforce and monitor the expected behaviors (compliant with MUD profile) of each IoT device, and train one-class classifier models to detect volumetric attacks such as DoS, reflective TCP/UDP/ICMP flooding, and ARP spoofing on IoT devices. 

Our \textbf{first} contribution develops a multi-level inferencing model to dynamically detect anomalous patterns in network activity of MUD-compliant traffic flows via SDN telemetry, followed by packet inspection of anomalous flows. This provides enhanced  fine-grained visibility into distributed and direct attacks, allowing us to precisely isolate volumetric attacks with microflow (5-tuple) resolution. For our \textbf{second} contribution, we collect traffic traces (benign and a variety of volumetric attacks) from network behavior of IoT devices in our lab, generate labeled datasets, and make them available to the public. Our \textbf{third} contribution prototypes a full working system (modules are released as open-source), demonstrates its efficacy in detecting volumetric attacks on several consumer IoT devices with high accuracy while maintaining low false positives, and provides insights into cost and performance of our system. Our \textbf{last} contribution demonstrates how our models scale in environments with a large number of connected IoTs (with datasets collected from a network of IP cameras in our university campus) by considering various training strategies (per device unit versus per device type), and balancing the accuracy of prediction against the cost of models in terms of size and training time.

\end{abstract}


\IEEEpeerreviewmaketitle
\vspace{-0.4cm}
\section{Introduction}\label{sec:intro}
The proliferation of insecure Internet-connected devices is making it easy \cite{loit2017systematically} for cyber-hackers to attack smart environments and infrastructures at large scale. Recent reports \cite{f5Labs17} show that attackers continue to exploit insecure IoT devices to launch volumetric attacks in the form of DoS, DDoS, brute force, and TCP SYN/UDP flooding. Moreover, the progression of botnets \cite{CiscoReport17, CiscoReport18} such as Mirai and Persirai, infecting millions of IoT devices, is enabling destructive cyber-campaigns of unprecedented magnitude to be launched. 


Today, IoT network operators are unable to verify whether their connected  IoT assets behave normally or not \cite{CiscoReport17}. In fact, most operators would not even know what ``normal'' behavior is, given the myriad IoT devices available with different functionalities and sourced from various manufacturers. To alleviate this issue, Manufacturer Usage Description (MUD) \cite{ietfMUD18} standard was proposed to alleviate this issue by requiring vendors to formally specify the intended network behavior of the IoT devices they make. The adoption of MUD specifications has been relatively slow. However, large organizations and critical infrastructure are increasingly deploying IoT devices at scale, and thus demand an automated enforcement of baseline security for multitude of IoT devices across their network. 

This new standard allows an operator to lock down the network behavior of the IoT device using access control lists (ACLs) derived from its MUD profile; indeed, our earlier work \cite{IoTSnP18-mudids} used software defined networking (SDN) to automatically enforce MUD profiles to the network -- 
translating ACLs into static and dynamic flow rules that can be applied at run-time on Openflow-capable switches to limit IoT traffic, thereby significantly reducing their  attack surface.  

The focus of this paper is on attacks that can be launched on IoT devices while still conforming to their MUD profiles. 
In particular, volumetric attacks on an IoT device are not necessarily prevented by its MUD profile, because its ACEs (access control entries) simply allow or deny traffic, and there is no provision to limit rates.
In this paper we show that a range of volumetric attacks (including ones directly on the IoT device and ones that reflect off the IoT device) are feasible in spite of MUD policy enforcement in the network. These volumetric attacks can be categorized into two types: (a) attacks like Flash crowd \cite{ari2004modeling} and Worms \cite{spafford1989internet} that are launched by generating a large and dynamic number of microflows (5-tuple), each at slow rates, resulting in a distributed attack (DDoS), and (b) attacks that are launched over a small and static number of microflows, each with a high volume of traffic (direct or point-to-point attack). Fending off such attacks requires a more sophisticated machinery that judiciously monitors volume and dispersion of microflows associated with each ACE rule to detect anomalies. 	Inspired by prior work \cite{TMC2019,TNSM20,IoTJ2020} showing that IoT devices display predictable traffic patterns (in use of network protocols and activity cycles), making it feasible to train machine learning models for detecting abnormal behavior, which is otherwise difficult for general-purpose computers that exhibit much wider diversity in network behavior \cite{sommer2010outside}.

A wide range of anomalous events may occur on a network, and hence network operators expect their monitoring tools and intrusion detection systems to provide them with explainable reasoning on detected anomalies (\cite{sommer2010outside}), \ie identifying attacks or unexpected behaviors at microflow levels. This would help operators mitigate cyber-attacks effectively, and possibly prevent (\eg black-listing external malicious entities) such attacks in future. Therefore, this paper aims to develop a method that not only detects anomalous IoT devices on a network but also identifies anomalous microflows for isolation/mitigation of attack and also better explaining anomalies and/or attacks. Our specific contributions are as follows:

\textbf{First}, we develop a system that learns benign behaviors of each IoT device by monitoring the activity patterns of its MUD-compliant traffic at various time scales via a combination of coarse-grained (per-device), medium-grained (per-service), and fine-grained (per-microflow) SDN telemetry. This system  is able to detect volumetric attacks (anomalous behavior) and identify specific traffic streams that contribute to the attacks. 
\textbf{Second}, we measure and record network behavior of real IoT devices under normal and attack conditions in our lab -- we subject our devices to volumetric attacks including ARP spoof, TCP SYN flooding, Fraggle, Ping of Death, and SSDP/SNMP/TCP /ICMP reflection. We label our traffic traces (\ie benign and attack), and make our data openly available to the research community.
\textbf{Third}, we prototype our system using an OpenFlow switch, Faucet SDN controller, and a MUD policy engine, and quantify its efficacy in detecting volumetric attacks on several IoT devices and isolating microflows which contribute to the attack. We release our solution modules as open-source to the community. \textbf{Lastly,} we demonstrate the scalability of our machine learning models by using a dataset collected from university campus and home networks, and draw insights into the accuracy of detection, the memory footprint of models, and training time depending upon various strategies such as modeling IoT device instance versus IoT type.


\section{Related Work} \label{sec:prior}

Intrusion detection systems have been studied extensively by the research community, and are typically in the form of checking for \textit{signatures} of known attacks (``blacklisting''), \textit{anomalies} indicative of deviation from normal behavior, or \textit{specification} of allowed traffic (``whitelisting'').
However, there are limited studies on detecting intrusions for IoT devices \cite{mlddosiot2018}. Security of IoT devices is increasingly becoming important due to their limited protection, if any.

\textbf{Signatures-based intrusion detection:} 
Nearly all deployed solutions, including software tools like Bro\cite{Bro1999} and Snort \cite{Roesch1999}, and commercial hardware appliances belong to this category. 
There are studies that apply signature-based intrusion detection and/or prevention in SDN environments \cite{Yoon2015, Piggybacking17}.
Signature-based approach is not sufficient for addressing the new and growing security issues that come with the proliferation of IoT devices. Attack signatures can not be developed for a growing number of IoT devices at scale. We will show (in \S\ref{sec:comparison}) that signature-based tools are only able to detect limited number of attacks (on IoT devices) those that are common for general purpose computers.

\textbf{Anomaly-based intrusion detection:} 
Anomaly detection holds promise as a way of detecting new and unknown threats, but despite extensive academic research \cite{Garcia2009}, has had very limited success in operational environments. The reasons for this are manifold \cite{sommer2010outside}: ``normal'' network traffic can exhibit much more diversity than expected (particularly for general-purpose devices); obtaining ``ground truth'' on attacks in order to train the classifiers is difficult; evaluating outputs can be difficult due to the lack of appropriate datasets; false positives incur a high cost on network administrators to investigate; and there is often a semantic gap between detection of an anomaly and actionable reports for the network operator.
There are many studies that employ either entropy-based \cite{lakhina2005mining, kumar2007distributed, mehdi2011revisiting,giotis2014combining} or machine learning \cite{braga2010lightweight,cui2016sd,tang2016deep} techniques to detect new attacks in SDN environments.
Entropy-based approaches are used for detecting types of attacks that generate a large number of flows. Authors in \cite{lakhina2005mining, kumar2007distributed,giotis2014combining} use sample entropy of source/destination IP address and port number to determine whether significant variation is observed by checking the measured entropy against a threshold. Works in \cite{kumar2007distributed}, \cite{lakhina2005mining} and \cite{giotis2014combining}  apply this technique to detect attacks in an ISP network, a backbone network, and a campus network respectively. The entropy check has proven to be effective in detecting anomalies, but it does not provide information on the cause of attacks. We note that identifying attack mircroflows in networks with high bit-rates can be quite challenging due to cost of processing. In this paper, we develop and demonstrate a dynamic and cost-effective method to identify microflows in volumetric attacks on IoT devices.
Some of prior works \cite{braga2010lightweight,cui2016sd,tang2016deep} employ binary (\ie benign and attack) classifiers. 
However, this approach contradicts with the expectation from anomaly-based techniques that need to flag deviations from normal behaviors \cite{sommer2010outside}. 
Authors of \cite{braga2010lightweight, cui2016sd} propose to use features including flow-level stats (\ie packet/byte count and duration), percentage of bidirectional flows, growth rate of unidirectional flows, and growth rate of number of unique ports, for their classifier. Work in \cite{tang2016deep} employs deep learning algorithms using a similar set of features to classify normal and abnormal traffic. Authors of \cite{bhunia2017dynamic} applied a techniques in  \cite{braga2010lightweight} to IoT devices. However, their evaluation is limited to simulated traffic in mininet that does not represent behavior of real IoT devices. 

\textbf{Specification-based intrusion detection:} Specifying allowed rules for general-purpose devices has been a challenge \cite{specification2001} as traffic pattern highly depend on applications and user activity. In \cite{jamaral}, the authors propose a specification-based approach for a wireless sensor network, and expect the network operator to define the rules. We believe this is too onerous for the network operator; the behavior is better defined by the manufacturer of the IoT device, which is exactly what IETF's MUD proposal \cite{ietfMUD18} intends. Our work in \cite{IoTSnP18-mudids} is the first that proposes an IDS for IoT devices using a combination of MUD and SDN and detects attack flows that are not specified in the device formal MUD profile. 
This work employs a collection of anomaly workers each trained by MUD behavioral profile that detect attacks that conform with MUD profile but display a deviated traffic profile.

\vspace{-0.2cm}
\section{Anomaly Detection using SDN}\label{sec:design}
In this section we describe our attack detection solution, including a brief summary of MUD profile (\S\ref{sec:mud}), the SDN-based system architecture (\S\ref{sec:sdn}), the anomaly detector (\S\ref{sec:anomaly}).

\vspace{-4mm}
\subsection{MUD Profile}\label{sec:mud}
MUD is a relatively new standard \cite{ietfMUD18}. A valid MUD profile contains a root object called ``access-lists'' container that comprises several access control entries (ACE), serialized in JSON format. Access-lists are explicit in describing the direction of communication, \ie \textit{from-device} and \textit{to-device}. Each ACE would match on source/destination port numbers for TCP/UDP, and type and code for ICMP. The MUD specifications also distinguish \textit{local-networks} traffic from \textit{Internet} communications. 
The MUD standard defines how a MUD profile needs to be fetched and how the behavior of an IoT device needs to be defined. The MUD profile will be downloaded using a MUD url (\eg via DHCP option). 
IoT device manufacturers have not yet provided MUD profiles for their devices. But, 
we released the MUD profiles (automatically generated from packet traces) for 28 consumer IoT devices \cite{IoTSnP18-mudgee} -- in this paper, we use a subset of those profiles corresponding to devices that we experiment with. 
\begin{figure}[t!]
	\centering
	\includegraphics[width=0.45\textwidth]{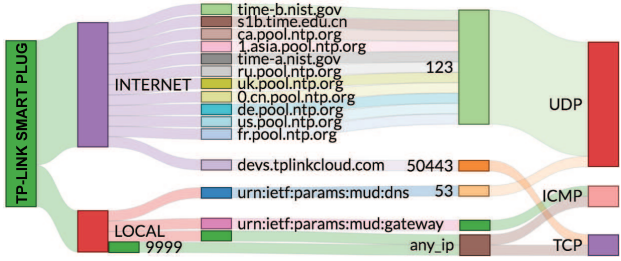}
	\vspace{-3mm}
	\caption{Sankey diagram showing MUD profile of TP-Link smart plug.}
	\label{fig:tplinksankey}
	\vspace{-4mm}
\end{figure}		

Fig.~\ref{fig:tplinksankey} visualizes a sample MUD profile in a human-friendly way, using a Sankey diagram to represent the MUD profile of a TP-Link smart plug. 
It is seen that this IoT device exchanges DNS queries/responses with the local DNS server, communicates with a range of Internet domains for NTP services (\ie UDP port 123), and talks to its manufacturer server (\ie {\myverb{devs.tplinkcloud.com}}) over TCP port 50443. In addition, the TP-Link smart plug exposes TCP port 9999 on the local network to its mobile app for user interaction with the device. 
It is also apparent that the smart plug and its mobile app send periodic pings to the gateway and the plug respectively for connectivity checking.


\subsection{SDN-Based System Architecture}\label{sec:sdn}

An IoT device advertises its MUD profile through a MUD URL. According to the MUD standard, there are three options for emitting the MUD URL namely DHCP, LLDP, and X.509 \cite{ietfMUD18}. If a device is compromised, the MUD URL emitted can potentially be spoofed in case of either DHCP or LLDP. But, it is secure when the device uses the X.509 extension since the MUD URL is added to the certificate by the manufacturer. This means that the MUD URL emitted by an X.509 device can not be spoofed without detection, even if the device is exploited.

\begin{figure}[t!]
	\centering
	\includegraphics[width=0.98\linewidth, height=4.5cm]{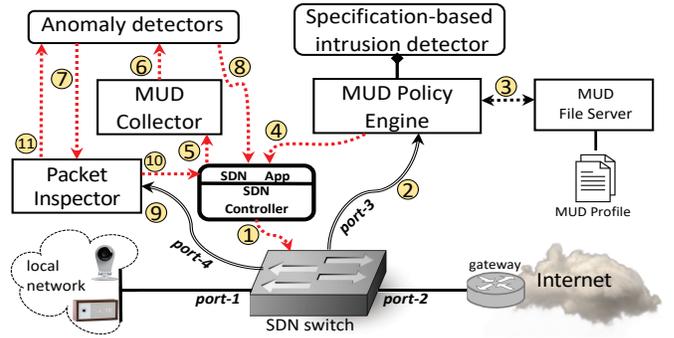}
	\vspace{-3mm}	
	\caption{Architecture of our SDN-based intrusion detection system.}
	\label{fig:systemarchitecture}
	\vspace{-4mm}
\end{figure}

Fig.~\ref{fig:systemarchitecture} shows the functional blocks in our architecture applied to a typical home or enterprise network. IoT devices on the left can communicate with other devices on the local network via a switch and also with Internet servers (not shown) via a gateway. 
In this architecture, the switch is an SDN switch whose flow-table rules will be managed dynamically by an external SDN controller. Our system includes a MUD policy engine in conjunction with an SDN App (on top of the controller), a MUD collector, and a combination of anomaly-based and specification-based threat detectors. 
The SDN App interacts with the MUD policy engine \cite{IoTSnP18-mudids} to insert proactive flow entries (for ACEs with known endpoints), and reactive flow entries (based on run-time DNS bindings) into the SDN switch. The App also interacts with the MUD collector to periodically pull flow volume data in the form of flow counters from the SDN switch. 
These components interact with each other to dynamically manage the flow-table rules inside the switch while monitoring the network activity of various flows pertinent to each device.

Note that the SDN switch does not redirect data packets to the SDN controller; rather, packets that need to be inspected are sent as copies on a separate interface of the switch, to which a software inspection engine is attached, as described below. This protects the SDN controller from overload from the data-plane, allowing it to scale to high rates and to service other SDN applications. Moreover, since incoming data packets are sent onwards by the switch immediately, the data-plane benefits in having minimal latency overhead, and is protected from failures of the SDN controller.


\definecolor{electricyellow}{rgb}{1.0, 0.8, 0.5}

\begin{table*}
	\centering
	\caption{Service flow rules for TP-Link Smart Plug.}\label{table:devrules}
	\vspace{-3mm}
	\begin{adjustbox}{max width=0.75\textwidth}	
		\renewcommand{\arraystretch}{1.2}
		\begin{tabular}{|l|c|c|c|c|c|c|c|c|c|c|}
			\hline 
			\textbf{flow-id} & \textbf{sEth} & \textbf{dEth} & \textbf{typeEth} & \textbf{Source} & \textbf{Destination} & \textbf{proto} & \textbf{sPort} & \textbf{dPort} & \textbf{priority} & \textbf{action}\tabularnewline
			\hline
			\rowcolor{electricyellow}
			a.1 & $<$ {\myverb{gwMAC}}$>$ & $<${\myverb{devMAC}}$>$  &  {\myverb{0x0800}} & [ntp domain names] & {\myverb{*}} & 17 & 123 & {\myverb{*}} & 20 & forward\tabularnewline
			\hline 
			\rowcolor{electricyellow}
			a.2 & $<${\myverb{devMAC}}$>$  & $<${\myverb{gwMAC}}$>$ &  {\myverb{0x0800}} & {*} &  [ntp domain names]  & 17 & {\myverb{*}} & 123 & 20 & forward\tabularnewline
			\hline
			\rowcolor{electricyellow}
			b.1 & $<${\myverb{gwMAC}}$>$ & $<${\myverb{devMAC}}$>$  &  {\myverb{0x0800}} & devs.tplinkcloud.com  & {\myverb{*}} & 6 & 50443 & {\myverb{*}} & 20 & forward\tabularnewline
			\hline 
			\rowcolor{electricyellow}
			b.2 & $<${\myverb{devMAC}}$>$  & $<${\myverb{gwMAC}}$>$ &  {\myverb{0x0800}} & {\myverb{*}} &  devs.tplinkcloud.com  & 6 & {\myverb{*}} & 50443 & 20 & forward\tabularnewline
			
			\hline
			c & $<${\myverb{devMAC}}$>$  & {\myverb{*}}  &   {\myverb{0x888e}} & {\myverb{*}} & {\myverb{*}} &{\myverb{*}} & {\myverb{*}} & {\myverb{*}} & 11 & forward \tabularnewline
			\hline
			d.1 & $<${\myverb{devMAC}}$>$  & ${\myverb{FF:FF:FF:FF:FF:FF}}$  &   {\myverb{0x0800}} & {\myverb{*}} & {\myverb{*}} &17 & {\myverb{*}} & 67 & 11 & forward \tabularnewline	
			\hline 
			d.2 & $<${\myverb{gwMAC}}$>$  & $<${\myverb{devMAC}}$>$   &  {\myverb{0x0800}} & {\myverb{*}} & {\myverb{*}} &17 & 67 & {\myverb{*}} & 11 & forward \tabularnewline			
			\hline 
			e.1 & $<${\myverb{gwMAC}}$>$ & $<${\myverb{devMAC}}$>$  & {\myverb{0x0800}} & gateway IP & {\myverb{*}} & 1 & {\myverb{*}} & {\myverb{*}} & 11 & forward\tabularnewline
			\hline
			e.2 & $<${\myverb{devMAC}}$>$ & $<${\myverb{gwMAC}}$>$  & {\myverb{0x0800}} & {\myverb{*}} & gateway IP& 1 & {\myverb{*}} & {\myverb{*}} & 11 & forward\tabularnewline
			\hline
			f.1 & $<${\myverb{devMAC}}$>$  & $<${\myverb{gwMAC}}$>$ &  {\myverb{0x0800}} & {\myverb{*}} &  gateway IP & 17 & * & 53 & 11 & forward\tabularnewline
			\hline
			f.2 & $<${\myverb{gwMAC}}$>$ & $<${\myverb{devMAC}}$>$  & {\myverb{0x0800}} & gateway IP & {\myverb{*}}  & 17 & 53 & {\myverb{*}} & 11 & forward \& mirror\tabularnewline
			\hline
			g.1 & $<${\myverb{devMAC}}$>$  & $<${\myverb{devMAC}}$>$ & {\myverb{0x0800}} & {\myverb{*}} & {\myverb{*}} & {\myverb{*}} & {\myverb{*}} & {\myverb{*}} & 10 & forward \& mirror\tabularnewline
			\hline  
			g.2 & $<${\myverb{gwMAC}}$>$ & $<${\myverb{devMAC}}$>$  &{\myverb{0x0800}} & {\myverb{*}} & {\myverb{*}} & {\myverb{*}} & {\myverb{*}} & {\myverb{*}} & 10 & forward \& mirror\tabularnewline
			\hline 
			h.1 & {\myverb{*}} & $<${\myverb{devMAC}}$>$  & {\myverb{0x0806}} & {\myverb{*}} & {\myverb{*}} & {\myverb{*}} & {\myverb{*}} & {\myverb{*}} & 7 & forward\tabularnewline
			\hline 
			h.2 & $<${\myverb{devMAC}}$>$  & {\myverb{*}} &  {\myverb{0x0806}}\verb|| & {\myverb{*}} & {\myverb{*}} & {\myverb{*}} & {\myverb{*}} & {\myverb{*}} & 7 & forward\tabularnewline
			\hline
			i.1 & $<${\myverb{devMAC}}$>$ & {\myverb{*}} & {\myverb{0x0800}} & {\myverb{*}} & {\myverb{*}} & 6 & 9999 & {\myverb{*}} & 6 & forward\tabularnewline
			\hline
			i.2 & {\myverb{*}} & $<${\myverb{devMAC}}$>$  & {\myverb{0x0800}} & {\myverb{*}} & {\myverb{*}} & 6 & {\myverb{*}} & 9999 & 6 & forward\tabularnewline
			\hline
			j.1 & $<${\myverb{devMAC}}$>$ & {\myverb{*}}  & {\myverb{0x0800}}& {\myverb{*}} & {\myverb{*}} & 1 & {\myverb{*}} & {\myverb{*}} & 6 & forward\tabularnewline
			\hline
			j.2 & {\myverb{*}} & $<${\myverb{devMAC}}$>$  & {\myverb{0x0800}} & {\myverb{*}} & {\myverb{*}} & 1 & {\myverb{*}} & {\myverb{*}} & 6 & forward\tabularnewline
			\hline
			k & {\myverb{*}} & $<${\myverb{devMAC}}$>$  & {\myverb{0x0800}} & {\myverb{*}} & {\myverb{*}} & {\myverb{*}} & {\myverb{*}} & {\myverb{*}} & 5 & forward \& mirror\tabularnewline
			\hline 
		\end{tabular}
	\end{adjustbox}
	\vspace{-4mm}
\end{table*}

The operational flow of events in Fig.~\ref{fig:systemarchitecture} is as follows: the switch is initially configured by a default rule, as shown by step \ciao{1}, to mirror  packets (on port-3), as shown by step \ciao{2}, that reveal the device identity (\eg DHCP traffic), and all other packets are forwarded normally (on either port-1 or port-2 depending on local or Internet communications). We note that DHCP packets contain the MAC of the device, and may also provide a \textit{mud-url} if the corresponding device manufacturer adopts the MUD standard. This assists the MUD policy engine to discover a new IoT device connected to the network -- the MUD engine keeps track of already discovered devices. 
Thereafter, the MUD engine fetches the corresponding MUD profile from the MUD file server, as shown by step \ciao{3} -- The MUD engine stores the fetched profile until its validity period expires. 
In a real scenario, the MUD file server will be operated by the manufacturer who can update the device MUD profile when needed (\eg due to a firmware upgrade).

The MUD policy engine translates access control entries (ACEs) of the MUD profile into a set of flow rules (explained in \S\ref{sec:anomaly}) we call them ``MUD flows'' or ``service flows'' that match on the 3-tuple or 4-tuple. MUD specifications allow manufacturers to specify Internet endpoints by their domain-name in ACEs. These ACEs can not be directly translated to flow rules and need further inspection to infer DNS bindings. The MUD engine, therefore, inserts proactive flow entries, shown by step \ciao{4}, for ACEs with known endpoints (\ie static IPs) while others are reactively inserted based on run-time DNS bindings. 
An idle-timeout is set for reactive flow rules that are associated with a domain name, to account for dynamic DNS bindings.

Following insertion of device flow rules, the switch mirrors all DNS responses in addition to exception packets that do not match on any proactive or reactive flow rule (\ie default mirror of local and Internet traffic). 
These mirrored packets are inspected by a ``specification-based intrusion detector'' component of the MUD policy engine to detect traffic that does not conform to the MUD profile of the corresponding IoT device.
The specification-based intrusion detector maintains an intermediate set of rules translated from the MUD profile, along with a DNS cache (all in memory) to determine whether headers of the mirrored packet match the intended profile of the corresponding IoT device.  Once an exception packet is matched to a DNS cache entry, a corresponding reactive flow rule is added to the flow rules of the SDN switch.

We note that a sophisticated attack traffic can still pass undetected \cite{IoTSnP18-mudids} using spoofing techniques, so that the attack traffic conforms to the MUD profile(s) of the IoT device(s) under attack, and will therefore not be detected by the specification-based intrusion detector.
n order to identify such ``volumetric'' attack threats, our system monitors the activity of all device flows specified by the MUD profile. To do so, the MUD collector periodically pulls flow counters denoted by step \ciao{5} in Fig.~\ref{fig:systemarchitecture}) from the switch, computes traffic volume attributes for each IoT device, and streams them to the corresponding anomaly detector for that IoT device, as denoted by step \ciao{6}.

We have two types of anomaly detectors: (a) volumetric-based detectors that monitor the activity volume (waveform) of individual MUD flows to identify anomalous behaviors, and (b) dispersion-based detectors that monitor the dispersion of microflows (associated with a MUD flow) to detect distributed attacks.
Once the activity volume of a  MUD flow (service flow) is found to be out of its norm, the corresponding model sends two signals: one to the  packet inspector module with identity of the anomalous MUD flow (step \ciao{7}), and another to the SDN App to install a reactive flow (step \ciao{8}) for mirroring (step \ciao{9}) all packets that match the anomalous MUD flow to port-4 -- the packet inspector is expected to check dispersion of microflow(s) from mirrored packets of the anomalous MUD flow.

Next, the packet inspector module performs two tasks: firstly, it reactively inserts microflows of highest priority, matching 5-tuple, into the switch (via SDN App by step \ciao{10}) in order to stop mirroring of packets that have been identified; secondly, it computes dispersion features (explained in \S\ref{sec:featureExtract}) and calls a corresponding anomaly detector to check whether the collection of microflows displays a distributed attack, or not -- this is indicated by step \ciao{11}.
In parallel, volumetric anomaly detectors continue to monitor the volume of these microflows, checking for the presence of any direct attacks. These microflows automatically	time out upon a duration of inactivity, so as to reduce TCAM usage -- the idle timeout value needs to be determined statically by network administrator and/or dynamically based on current entries inserted into the switch. Also, note that reactive microflows provide fine-grained telemetry for detecting the attack flow(s). In what follows we explain our features and anomaly detection algorithm.

\vspace{-0.3cm}	
\subsection{Anomaly Detection Method}\label{sec:anomaly}
We develop a machine learning technique (explained in \S\ref{sec:anomalydetectionworkers}) to determine if an IoT device is involved in a volumetric attack or not
(the ``attack detection''), and if it is, to identify the microflow that contributes to the attack (the ``attack flow(s) identification'').
Our objective is to train models with benign traffic profile of each device, and detect attacks by detecting deviations from expected traffic pattern of a device flows defined by the device MUD profile.

\subsubsection{Device Flow Rules}

For our anomaly detection, we consider two types of flow rules: (a) MUD flows (service flows) that are 3-tuple or 4-tuple rules derived from the MUD profile of an IoT device, and (b) microflows (corresponding to a MUD flow) that are 5-tuple rules identified by the packet inspector module.

\begin{table*}
	\centering
	\caption{Microflow rules for TP-Link Smart Plug.}\label{table:stage3attackflows}
	\vspace{-4mm}
	\begin{adjustbox}{max width=0.85\textwidth}	
		\renewcommand{\arraystretch}{1.2}
		\begin{tabular}{|l|c|c|c|c|c|c|c|c|c|c|}
			\hline 
			\textbf{flow-id} & \textbf{sEth} & \textbf{dEth} & \textbf{typeEth} & \textbf{Source IP} & \textbf{Destination IP} & \textbf{proto} & \textbf{sPort} & \textbf{dPort} & \textbf{priority} & \textbf{action}\tabularnewline
			\hline
			
			i.1- reactive & $<${\myverb{devMAC}}$>$ & *  & {\myverb{0x0800}}& {\myverb{192.168.1.227}} & {\myverb{192.168.1.228}} & 6 & 9999 & 43847 & 30 & forward\tabularnewline
			\hline
			\hline
			\rowcolor{electricyellow}
			i.1 & $<${\myverb{devMAC}}$>$ & *  & {\myverb{0x0800}} & {\myverb{devIP}} &  {\myverb{*}} & 6 & 9999 & {\myverb{*}} & 6 & forward \& mirror\tabularnewline
			\hline
		\end{tabular}
	\end{adjustbox}
	\vspace{-5mm}
\end{table*}

\textbf{MUD service flow rules:}
As briefly explained in \S\ref{sec:sdn}, a given MUD profile is processed to generate corresponding flow-table rules \cite{IoTSnP18-mudgee} that are used to monitor the expected traffic of the device. 
For example, in \autoref{table:devrules}, we show service flow rules generated from the MUD profile of a TP-Link smart plug IoT device. 
The highlighted rows (\ie flow-IDs \textit{a.1}, \textit{a.2}, \textit{b.1} \& \textit{b.2}) correspond to a snapshot of reactive flow rules that may vary over time. 
Reactive rules have a priority slightly higher than of default flows mirroring traffic. This way, we stop mirroring packets of traffic flows that conform to the MUD profile. 
In this table, Internet sources/destinations are shown by domain-names to make it easier to visualize (IP addresses are used in the actual flow-table). The un-highlighted rows correspond to proactive rules. Proactive rules \textit{f.2}, \textit{g.1 \& g.2}, and \textit{k}, respectively mirror: DNS replies, default Internet traffic from/to, and the local traffic to this device.
Only one direction of local traffic (\ie to the IoT device) is used to avoid conflicting with matching flows of other devices. Mirroring traffic coming to the device allows our system to inspect any attempt to access standard vulnerable services such as Telnet, SSH, or HTTP that might be open on IoT devices.

\textbf{Microflow rules:} Once an anomaly is detected in a MUD flow, all packets of that flow are mirrored for further header inspection (\ie a shallow inspection). The packet inspector identifies microflows by receiving a packet from them, and consequently inserts a reactive rule with the idle timeout equals to a minute, stopping subsequent packets of the identified microflows to be mirrored. Table~\ref{table:stage3attackflows} illustrates a partial snapshot of the switch flow table related to the TP-Link plug when its service worker ``\textbf{i}'' detects an anomalous behavior in the local communications, initiated from TCP port 9999. We can see that the original ``forward'' action (flow-id ``\textbf{i.1}'' in Table~\ref{table:devrules}) changes to  ``forward \& mirror'' (second row in Table~\ref{table:stage3attackflows}).
Note that mirroring traffic of a volumetric attack may paralyze the packet inspector module, however packet-forwarding remains unaffected and operational. In order to protect the packet inspector, one can apply rate-limiting on the egress interface of the mirror port. Also, TCAM usage can be managed dynamically based on the total capacity and growth rate of flow entries. These protection mechanisms are beyond the scope of this paper.



\subsubsection{Attack Detection Models}
\begin{figure}[t!]
	\centering
	\includegraphics[width=0.95\linewidth]{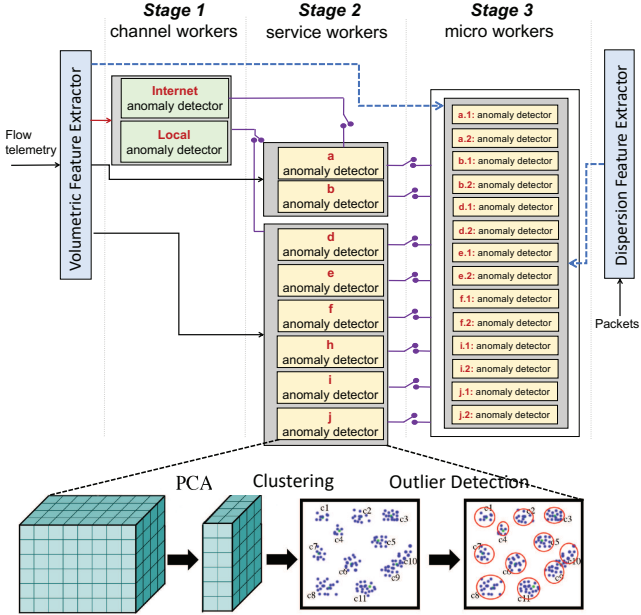}
	\vspace{-3mm}
	\caption{Structure of anomaly detection models for TP-Link smart plug.}
	\label{fig:attackdetector}
	\vspace{-4mm}
\end{figure}

We now present our three-stage approach for detecting anomalous flows. For each device, a specialized set of models are trained based on network activities of its MUD service flows and microflows. 
The process is not only able to detect an attack on the device, but also to identify the microflow(s) contributing to the attack.
For example, Fig.~\ref{fig:attackdetector} depicts the structure of models specific to the TP-Link smart plug for detecting anomalies caused by volumetric attack traffic. 
The process first identifies whether an anomaly occurs over local or Internet communication using respective separate one-class device-specific classifiers referred to herein as ``channel detectors'' or ``workers'' (stage-1) -- these workers utilize coarse-grained (device-level) telemetry. A true alarm from the stage-1 workers triggers corresponding one-class flow classifiers (also referred to herein as ``detectors/workers'') at stage-2 which identify the MUD flow(s) over which the attacker causes the anomaly using specialized service flow detectors/workers, each corresponding to a MUD service flow in \autoref{table:devrules} -- these workers utilize medium-grained (service-level) telemetry. 

We note that our inferencing at stage-1 and stage-2 is not able to distinguish ``direct'' volumetric attacks from ``distributed'' ones since network telemetry is (to some extent) aggregated at these two stages -- channel flows and service flows can be aggregate of several microflows. However, each type of volumetric attacks (direct or distributed) requires a specific strategy for an efficient mitigation, and hence a microscopic inferencing at stage-3 is essential. A true alarm from the stage-2 workers would trigger stage-3 inferencing whereby we mirror all packets that match the MUD flow classified as anomalous by its corresponding model. At stage-3, our objective is to determine whether the attack is distributed or direct, and thereby detect microflows which cause anomaly in the behavior of the flagged MUD flow. We, therefore, employ both volumetric-based and dispersion-based workers in parallel for inferencing at stage-3. 


Note that an attack is certainly detected only when all the three stages generate true alarms of anomaly for a given IoT device. Due to high cost of inferencing at stage-3, it is not practical to have all three stages to work in parallel for large-scale IoT networks, and hence short-duration attacks may get missed -- sophisticated short-duration attacks can be scheduled to rapidly emerge and vanish, to bypass intrusion detection systems. Instead, for a specific device, one can activate the stage-3 inferencing to operate in parallel to the other two stages when frequent alarms are generated by stage-1 and stage-2. In \S\ref{attackdetx} we will show how a combination of the three-stage inferencing allows us to reduce  the rate of false-positives and the cost of packet inspection, while maintaining a high rate true-positives.


\subsubsection{Features Extractor}\label{sec:featureExtract}
Having captured service flow rules for each IoT device (\eg \autoref{table:devrules} for the TP-Link smart plug), corresponding features of  network activity are then extracted.  
We extract two types of features: (a) flow counters to train volumetric-based anomaly detection, and (b) spread of packet headers to train dispersion-based anomaly detection. These two feature extractors are respectively fed by flow telemetry and mirrored packets, as shown in Fig.~\ref{fig:attackdetector}. Note that dispersion features are collected once the stage-3 inferencing is activated.

\begin{figure}[t!]
	\begin{center}
		\mbox{
			\subfigure[UDP port 53 downstream.]{
				{\includegraphics[width=0.2\textwidth]{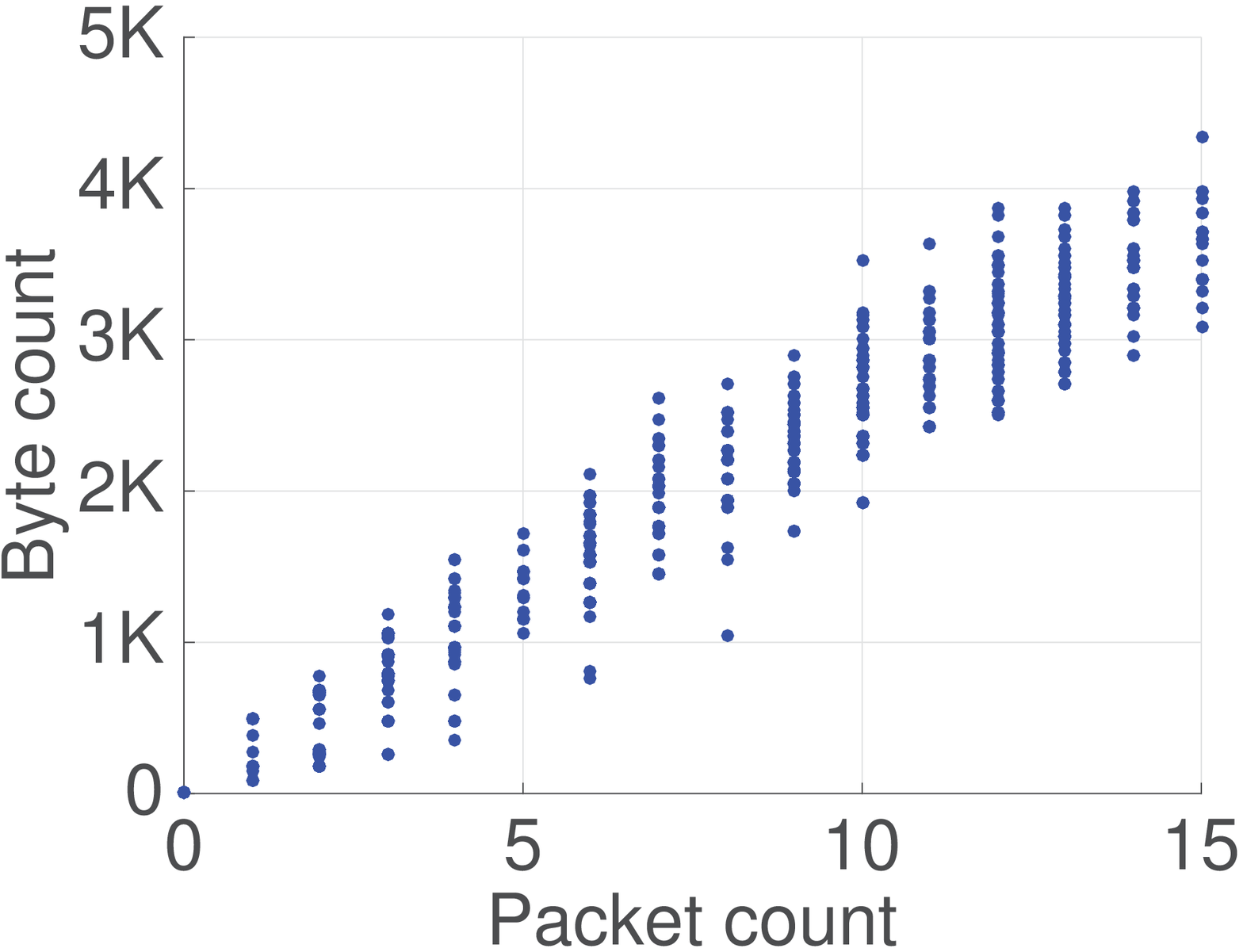}}\quad
				\label{fig:feature53}
			}
			\subfigure[TCP port 465 downstream.]{
				{\includegraphics[width=0.2\textwidth]{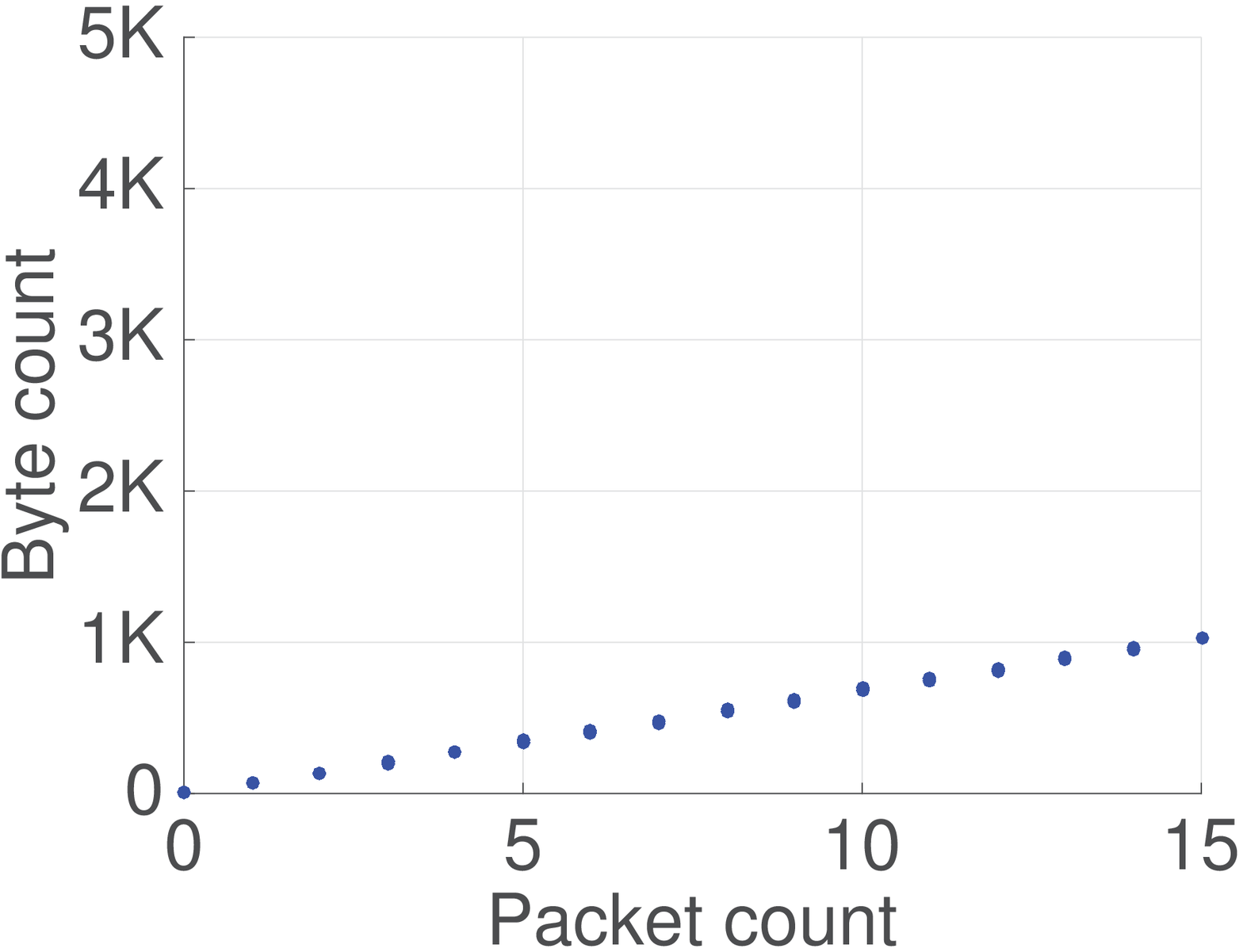}}\quad
				\label{fig:feature465}
			}
		}
		\vspace{-3mm}
		\caption{Byte count vs. packet count of downstream remote traffic to Samsung smart-cam.}
		\vspace{-7mm}
		\label{fig:featurerelation}
	\end{center}
\end{figure}

\textbf{Volumetric features:} We use the count of packets and bytes provided by each flow rule as volumetric features. This is because the size of packets can vary for a given service protocol. For example, Fig.~\ref{fig:feature53} shows the scatter plot of packet count versus byte count of DNS downstream traffic captured for Samsung camera over one month in our lab. It is seen that for a given packet count, the byte count varies over a range of 1 KB or more, , indicating that packet count and byte count are not highly correlated. However, for TCP port 465 downstream traffic for the same device, shown in Fig.~\ref{fig:feature465}, packet count and byte count are highly correlated (indicating a consistent packet size). 

Traffic features are also generated for multiple time-scales by retrieving flow counters (packet and byte counts) every minute and processing the counter values to generate values for the totals, means, and standard-deviations of packet and byte counts over sliding windows of, 2-, 3- and 4-minute (explained in \S\ref{sec:evaluation}) as features, and including the original byte and packet count values as an additional two features, providing a total of 20 features (also referred to herein as attributes) for each flow rule at any point in time. It is important to note that we only consider flow-level features (of static number of flows per device) for the first two stages of our inferencing to keep the computing cost at a minimum -- one may choose to use other sliding windows, features, and combinations.
Upon detection of attack service flow(s), we employ both packet-level and flow-level features (relatively expensive, but computed for a specific fraction of traffic of only anomalous devices) at stage-3 to identify the attack microflow(s).

For workers of the stage-1, we use attributes of a set of flows that share the channel specified by the MUD profile (\ie local or Internet) -- for example flows \textbf{a.1}, \textbf{a.2}, \textbf{b.1}, and \textbf{b.2} for the Internet channel. Each worker of the stage-2 corresponds to a bidirectional traffic flow (\ie a couple of flow rule to/from the device). For example, machine ``\textbf{a}'' of the stage 2 in Fig.~\ref{fig:attackdetector} uses features of two flows \textbf{a.1} and \textbf{a.2} from \autoref{table:devrules}.

There can be  multiple reactive rules for an Internet flow due to dynamic DNS bindings. We therefore aggregate these rules by wild-carding the Internet endpoint. It is important to note that default rules (\ie \textbf{g.1}, \textbf{g.2}, and \textbf{k}) are not considered for anomaly detection, as they are handled by the specification-based intrusion detector (explained in \S\ref{sec:sdn}). 

For volumetric-based anomaly detectors at stage-3, we use the same features used at stage-1 and stage-2 but computed for microflows. For example, if an anomaly is detected in service flow \textbf{a.1} then we identify and push its corresponding microflows with high priority into the switch (\S\ref{sec:sdn}), and compute volumetric features from the minutely counters of these microflows.

\textbf{Dispersion features:} 
An important trait of distributed attacks is the degree of dispersal or concentration across all the flows involved. For example, in a TCP distributed attack on a device, source IP and TCP port may vary, generating a large number of microflows, while the destination IP and TCP port are fixed and concentrated. To quantify the dispersion metric in a set of observations $X=\{x_{1},....,x_{N}\}$ where each observation $x_i$ occurs $n_i$ times, we use sample entropy \cite{richman2000physiological}, given by: 
\vspace{-3mm}


\begin{equation}\label{eq:eq1}
	{
		H(X) = - \sum_{i=1}^{N}p_{i} . log_{2}(p_{i})
	}
	\vspace{-2mm}
\end{equation}

where  $p_{i}=n_{i}/\sum n_i$ is the distribution of our observations. 
The entropy value $H(X)$ ranges from $0$ (concentrated) to $log_{2}(N)$ (diverse).

To determine the presence of a distributed volumetric attack on a service flow, we analyze the sample entropy of certain headers (\ie IP and transport layers) across all of its associated microflows. The entropy is only computed on those header fields (of microflows) that are wildcarded by the corresponding service flow. In this paper, we focus on header attributes including source IP address, destination IP address, source port, destination port, ICMP type, and ICMP code that are commonly used by IoT devices.

For example, TP-Link Smart Plug uses TCP/IP protocols over 14 service flows 
(\ie all flows except \textbf{g.1}, \textbf{g.2}, \textbf{k}, \textbf{h.1}, and \textbf{h.2}). Note that flows \textbf{g.1}, \textbf{g.2} and \textbf{k} are default rules, mirroring exception traffic that does not conform to the MUD profile, and \textbf{h.1}, and \textbf{h.2} correspond to ARP traffic. 
It should be noted that ARP spoofing attacks create a large number of ARP packets that can be flagged by the workers at both stage-1 and stage-2. However, determining whether those packets actually correspond to a spoofing attack requires payload inspection of ARP packets (\ie checking the mapping of IP address against MAC addresses) -- this specialized ARP worker is explained in \S\ref{attackflowiden}.

Note that distributed attacks can only be launched if they conform to MUD profiles, and hence they can only vary header fields that are wildcarded (dynamic) by a service flow. For example, the distributed attack shown in Fig.~\ref{fig:wormwemo}, concentrated on destination IP (the address of WeMo switch), protocol (TCP), and destination port number (49153), conforming to the MUD flow, while source IP and port number were dynamic. Therefore, for each dynamic header field a specialized anomaly detection model is needed to check the entropy of that specific header.  
To compute the entropy of microflow headers (streaming in real-time), we maintain a hash-map data structure of individual dynamic header fields over an epoch (say, every 5 seconds) -- unique observed values are counted during each epoch. Variation of entropy is captured by taking a sliding window of the entropy of individual features over successive epochs -- four epochs are considered to manage memory costs. This means that the model of each dynamic header field is trained by 4 features of entropy value.



\subsubsection{Anomaly Detection Workers} \label{sec:anomalydetectionworkers}

\begin{figure}[t!]
	\begin{center}
		\mbox{
			\subfigure[]{
				\includegraphics[width=0.26\textwidth,height=0.16\textwidth]{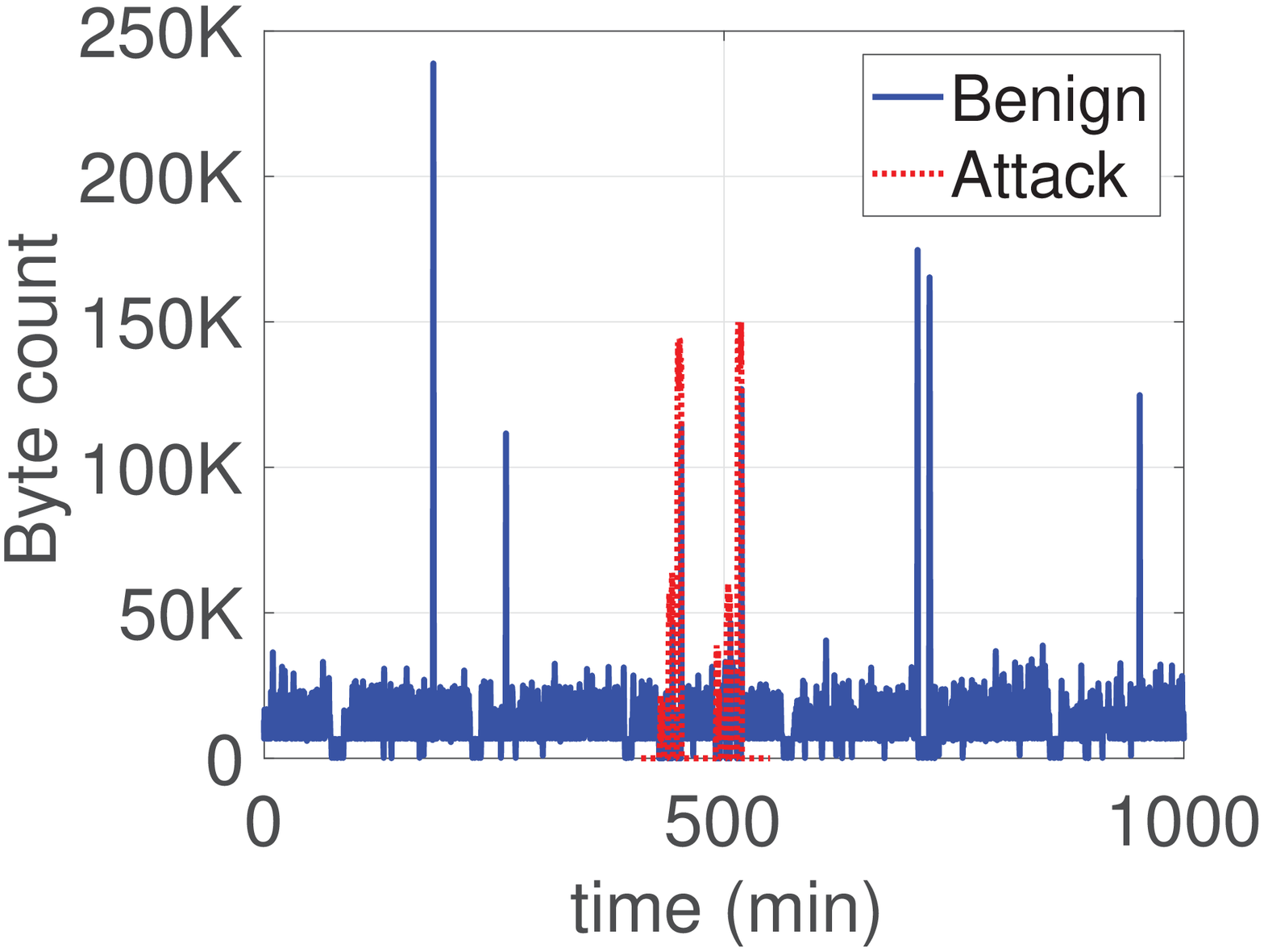}\quad
				\label{fig:netatmoCam}
			}
			\subfigure[]{\includegraphics[width=0.18\textwidth, height=0.15\textwidth]{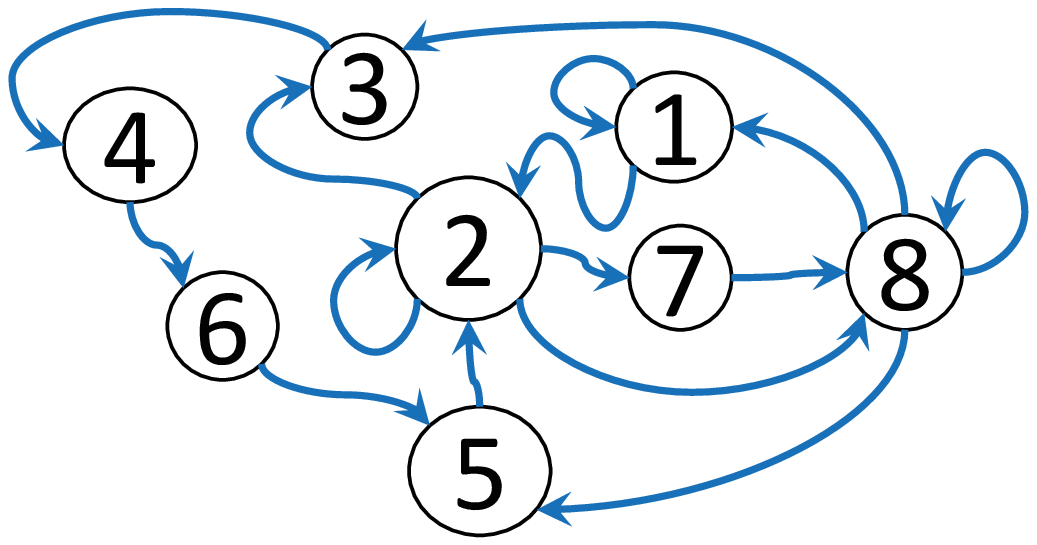}\quad
				\label{fig:transitionMatrix}
			}
		}
		\vspace{-5mm}
		\caption{(a) Traffic profile from an Internet server TCP port 443 to Netatmo camera. (b) State machine for the worker ``\textbf{a}'' of TP-Link plug.}
		\vspace{-7mm}
	\end{center}
\end{figure}

The anomaly detection workers (used in stage-1, stage-2 and stage-3) are based on the concept of one-class classification: device workers are trained by features of benign traffic of their corresponding IoT device, and are able to detect whether a traffic observation belongs to the trained class or not. Each anomaly detector uses a clustering-based outlier detection algorithm comprising three steps, as shown schematically at the bottom of Fig.~\ref{fig:attackdetector}, namely: (i) Principal Component Analysis (PCA), (ii) clustering, and (iii) boundary detection. 
Note that a simple thresholding would not be able to fully distinguish volumetric attack traffic from benign traffic. To better illustrate the case, we show in Fig.~\ref{fig:netatmoCam} profiles of benign traffic (solid blue lines) from an Internet server TCP 443 to Netatmo camera, and TCP SYN reflection attack (dashed red lines) to the same device. It is seen that no threshold value would detect the attack. This means that even having static rate-limits specified by the MUD profile may not be sufficient in detecting all attacks, instead it is needed to model and learn the dynamics of traffic activity behavior across various flows. 

\begin{figure}[t!]
	\centering
	\includegraphics[width=0.48\textwidth, height=0.27\textwidth]{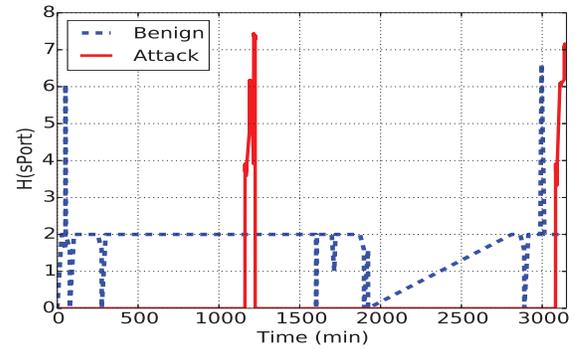}
	\vspace{-3mm}
	\caption{Timetrace of entropy for benign versus attack in traffic of WeMo switch over TCP port 49153.}
	\label{fig:wormwemo}
	\vspace{-5mm}
\end{figure}

A similar observation was made when attempting to set a threshold for accepted level of entropy in header fields of microflows. Fig.~\ref{fig:wormwemo} depicts a time-trace of entropy of source port during a benign communication over TCP port 49153 and an attack on a WeMo Switch. The attack was launched twice, each for a duration of 10 minutes. It can be seen that the entropy value of source port number during this attack fluctuates between 4 to 7, while the same entropy of the benign traffic varies from 0 to 6.5. Again, this clearly shows that having a single threshold value would not help us determine the spread of microflows, and hence a learning model (one-class classifier) of collective features extracted from microflows is required. 

For our one-class classification, we tried three main techniques \cite{pimentel2014review} namely probabilistic (\ie Gaussian mixture models), domain-based (\ie one-class SVM), and clustering-based (\ie DBSCAN, Kmeans), and found that the clustering approach performed the best in modeling the benign behavior of our IoT device types.



\textbf{Principle Component Analysis (PCA):} 
We note that each flow (service flow and microflow) contributes to 20 volumetric features -- the profile of the TP-Link power plug with 17 rules would result in a total of 340 features). This makes it computationally expensive to analyze such large number of features in real-time, specially for a large number of devices. However, there are features which are highly correlated (\eg Fig.~\ref{fig:feature465}) and can be transformed to reduce the feature space dimension. We, therefore, employ PCA \cite{abdi2010principal} to extract the principal components of our volumetric features that are orthogonal to each other -- we do not need to apply PCA to dispersion features since the cost of computing 4 features per model would be manageable, especially they only get activated when a service flow anomaly is detected.
We use Kaiser rule \cite{kaiser1960application} (eigenvalues >1) to deduce and select the most suitable set of principal components that capture all of the variance in the dataset.
As per the PCA requirement, all features are normalized using the z-scores method (\ie they are expressed as deviations from the mean divided by the standard deviation). 

\begin{table}
	\centering
	\caption{Anomaly detection model for TP-Link  plug.}
	\label{table:pcanalaysis}
	\vspace{-3mm}
	\begin{adjustbox}{max width=0.48\textwidth}	
		\renewcommand{\arraystretch}{1.2}
		\begin{tabular}{|l|c|c|c|c|c|c|c|c|c|c|}
			\hline 
			\textbf{Worker}& Local & Internet & a & b & d & e & f & h & i & j\tabularnewline
			\hline 
			\hline 
			\textbf{\# Features} & 261 & 81 & 41 & 41 & 41 & 41 & 41 & 41 & 41 & 21\tabularnewline
			\hline 
			\textbf{\# PCA} & 18 & 9 & 4 & 5 & 5 & 2 & 4 & 4 & 4 & 2\tabularnewline
			\hline 
			\textbf{Coverage (\%)} & 97.14 & 94.9 & 93.51 & 96.27 & 96.46 & 99.99 & 98.69 & 96.62 & 99.99 & 99.99\tabularnewline
			\hline 
			\textbf{\# Clusters} & 53 & 48 & \textbf{8} & 50 & 4 & 4 & 36 & 14 & 2 & 4\tabularnewline
			\hline 
		\end{tabular}
	\end{adjustbox}	
	\vspace{-6mm}
\end{table}

\textbf{Clustering:} As discussed earlier, we employ a number of anomaly detectors for each device, an efficient and computationally  inexpensive clustering algorithm is needed that: (a) can set the parameters automatically (\ie self-tuned), and (b) is able to deal with our benign dataset, containing a mix of sparse and dense regions. Among many possible clustering algorithms, we use X-means \cite{pelleg2000x} (\ie a variant of the K-means algorithm) that is a fairly lightweight yet efficient clustering method. The accuracy of high dimensional data clustering depends on the distance function used \cite{aggarwal2001surprising}. Conducting several experiments, Manhattan distance function provided us with the best results. Finally, we train the clustering algorithm with the principle components of our training dataset (obtained from PCA), providing as output the coordinates of the cluster heads, as shown by the brighter dots labeled by $c_{i}$ in Fig.~\ref{fig:attackdetector}.
%
Table~\ref{table:pcanalaysis} summarizes the count of features identified for each worker of the TP-Link plug. It is seen that how PCA reduces the feature dimension significantly while a high level of variations in the training dataset is covered. The last column shows the number of benign clusters created for each worker. 


\textbf{Outlier Detection:} We employ two outliers detection techniques (\ie boundary detection and state machine \cite{garcia2009anomaly}) to determine whether an instance is anomalous.

\textit{Boundary detection:} An anomaly is detected when an observation deviates from the clusters representing benign network traffic. 
Given the cluster heads and the training dataset, the 97.5th percentile is calculated as a boundary for each cluster, and anomalies observed outside these boundaries trigger an alarm, which is therefore expected to cause occasional mis-detections of benign traffic as anomalous (\ie false positive alarms).  


\textit{State Machine:} This technique flags anomalous instances that belong to one of expected clusters but their sequence of transition from the previous cluster (\ie state) is not normal. For this technique, we develop a state machine for each worker, capturing states transition across normal clusters. Fig~\ref{fig:transitionMatrix} illustrates an example of the state machine for a worker of TP-Link smart plug -- these 8 states correspond to clusters of worker ``\textbf{a}'' in Table~\ref{table:pcanalaysis}. Any transition outside of this chain will be raised as anomaly.

\subsection{Attack Microflow Identification}\label{attackflowiden}
In this subsection, we develop a method for stage-3 inferencing, to detect and isolate those microflow(s) that contribute to a volumetric attack, launched over a MUD flow. For stage-3, as explained in \S\ref {sec:anomaly}, two types of models get activated in parallel: (i) a specialized volumetric worker monitors the counters of individual reactive microflows, and (ii)  a set of dispersion workers monitor dynamic headers of microflows.

\textbf{Direct attacks} often use only a few and static number of microflows, and hence can be identified by the volumetric worker -- attack microflows can be precisely mitigated by changing their action to block (\eg via SDN). 
Note that there is an exception for handling anomalous ARP service flows (shown in \S\ref{attackdetx}) which can go out of their norm during  many volumetric attacks (ARP spoofing attack included) -- the actual cause of anomaly in ARP flows can only be determined by inspection of ARP packets payload. 
Therefore, we developed a specific ARP spoofing detector which gets activated when our stage-2 inferencing flags an anomaly in the ARP flow and  hence ARP packets are mirrored to packet inspector module. We extract the mapping of IP address to MAC from the payload of ARP packets (query and response), and look them up in the table that is progressively developed and updated by inspecting DHCP packets  -- recall from \S\ref{sec:sdn} that a default rule in the switch flow-table mirrors all DHCP packets. We verify (by look up) whether an IP address maps to more than one MAC address, or not -- the multiple mapped IP address is the identify of ARP spoofer, otherwise the anomaly in ARP flow is a side-effect of other attacks.  



\begin{figure}[t!]
	\centering
	\includegraphics[width=0.47\textwidth]{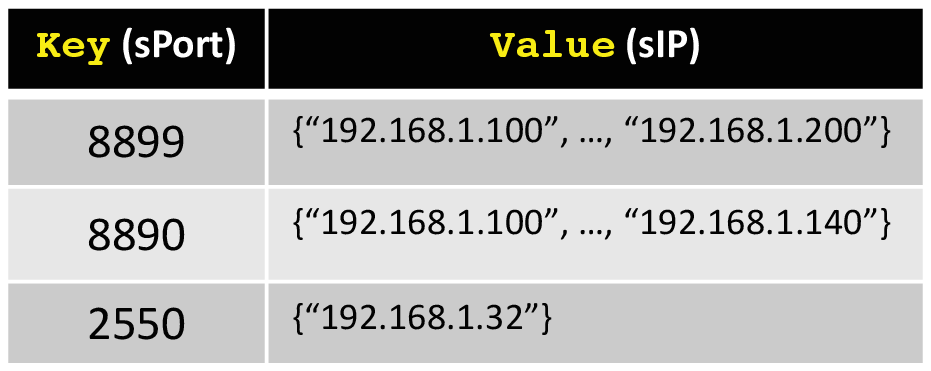}
	\vspace{-3mm}
	\caption{Hash-map data structure dynamic headers to identify maximum-matching mircroflows of a distributed attack.}
	\label{fig:hashMap}
	\vspace{-4mm}
\end{figure}

In \textbf{distributed attacks}, on the other hand, a large and dynamic number of microflows are generated, and hence reactive insertion of microflows can lead to TCAM exhaustion. 
Therefore, it is crucial to determine whether the anomaly in MUD flows is due to a distributed attack, or not. This task is performed (periodically in every epoch) by the set of dispersion-based models.
Once the attack is indicated (by at least one dispersion worker) to be distributed, all  reactive microflows (inserted into the switch) are removed to avoid TCAM exhaustion. It becomes impossible to identify attack microflows, if all dispersion workers indicative anomaly -- this happens when the attack is highly distributed across all dynamic headers, and thus the MUD flow itself is declared as the attack flow (more granular identification cannot be achieved).

Instead, when at least one dispersion worker indicates a benign output, we can consider a maximum-matching method across headers of the involved microflows. This means that we wildcard those dynamic headers whose dispersion model indicates a fast/anomalous change (\ie high entropy), and use only slow-changing headers to identify the attack flows. 

We determine the exact values of slow-changing headers (\ie low entropy and benign ones) by constructing a hash-map data structure.  In this map, the ``key'' of each entry is the value of slow-changing header -- it becomes a combined key in case of more than one slow-changing headers. The ``value'' of entries is a set, consisting of the values of fast-changing headers. 
Fig.~\ref{fig:hashMap} shows an example of the hash-map data structure for a distributed attack on the WeMo switch.  In this example, we launched the attack on the WeMo switch by generating a dynamic and growing number of connection requests to the device TCP port 49153, sourced from spoofed IP addresses.  It can be seen that the source IP addresses change very rapidly ($\sim$100 times per second), while only two source port numbers were repeatedly used by  varying IP addresses. We choose the key with the largest number of elements in its value set (\eg top row in Fig.~\ref{fig:hashMap}), identifying a group of distributed attack microflows. Once this group is mitigated, the identification process (using the hash-map) repeats in following epochs till no further anomaly is flagged by dispersion workers -- sophisticated attacks can only be precisely identified and mitigated in a progressive manner.  

\vspace{-3mm}
\section{Attack Tool and Data Collection}\label{sec:experiments}
\begin{table*}[!t]
	\begin{minipage}{.72\linewidth}
		\caption{Attacks launched on the IoT devices. (\textit{L:local, d:device, I:Internet})}
		\label{table:attacsklist}
		\centering
		\vspace{-3mm}
		\begin{adjustbox}{max width=0.98\textwidth}	
			\renewcommand{\arraystretch}{1.2}
			\begin{tabular}{|c|c||c|c|c||c|c|c|c|c|c|c|c|c|c||c|c|c|c|c|}
				\hline 
				\multicolumn{2}{|c||}{} & \multicolumn{3}{c||}{Maximum packet rate} & \multicolumn{10}{c||}{Device label} & \multicolumn{5}{c|}{Attack scenario}\tabularnewline
				\hline 
				\multicolumn{2}{|c||}{Attacks} & 1 pps & 10 pps & 100 pps & WM & WS & SC & TP & NC & CU & AE & PH & IH & LX &
				L$\rightarrow$d & L$\rightarrow$d$\rightarrow$L & L$\rightarrow$d$\rightarrow$I & I$\rightarrow$d$\rightarrow$I & I$\rightarrow$d\tabularnewline
				\hline 
				\multirow{4}{*}{Reflection} & SNMP & \cmark & \cmark & \cmark &  &  & \cmark &  &  & & & & & &  & \cmark & \cmark & \cmark & \tabularnewline
				\cline{2-20} 
				& SSDP & \cmark & \cmark & \cmark & \cmark &  &  &  &  & \cmark& & \cmark & & &  & \cmark & \cmark & \cmark &  \tabularnewline
				\cline{2-20} 
				& TCP SYN & \cmark & \cmark & \cmark & \cmark & \cmark & \cmark & \cmark & \cmark& \cmark& & \cmark & & &  & \cmark &  &\cmark  & \tabularnewline
				\cline{2-20} 
				& Smurf & \cmark & \cmark & \cmark &  & & \cmark & \cmark & & & & \cmark & & \cmark&  & \cmark &  &  &  \tabularnewline
				\hline 
				\multirow{5}{*}{Direct} & TCP SYN & \cmark & \cmark & \cmark & \cmark & \cmark & \cmark & \cmark & \cmark & \cmark& & \cmark & && \cmark &  &  &  &  \cmark\tabularnewline
				\cline{2-20} 
				& Fraggle & \cmark & \cmark & \cmark &  &  & \cmark &  &  &  & \cmark& &  & \cmark & &  &  &  & \cmark\tabularnewline
				\cline{2-20} 
				& Fraggle & \cmark & \cmark & \cmark &  \cmark   &  & \cmark &  &  &  & \cmark& &  & \cmark &\cmark &  &  &  & \tabularnewline
				\cline{2-20} 
				& Ping of Death & \cmark & \cmark & \cmark & \cmark & \cmark & \cmark & \cmark &  & & & \cmark & & \cmark& \cmark &  &  &  &  \tabularnewline
				\cline{2-20} 
				&  ARP Spoof &  \cmark  &  \cmark &  \cmark  & \cmark & \cmark & \cmark & \cmark & \cmark & \cmark  &  \cmark &  \cmark  & \cmark & \cmark & \cmark &  &  &  &  \tabularnewline
				\hline 
			\end{tabular}
		\end{adjustbox}
	\end{minipage}%
	\begin{minipage}{.26\linewidth}
		\centering
		\caption{Size of our  dataset.\label{table:summaryexperiment}}
		\vspace{-3mm}
		\begin{adjustbox}{max width=0.99\textwidth}	
			\renewcommand{\arraystretch}{1.2}			
			\begin{tabular}{|l|p{1.35cm}|p{1.2cm}|p{1.1cm}|p{1.0cm}|}
				\hline 
				Device  & \# train inst(min)& \# test inst(min)  & \# attack inst(min) & Device label\tabularnewline
				\hline 
				WeMo motion & 15000 & 61900 & 480 & WM\tabularnewline
				\hline
				WeMo switch & 15000 & 61861 & 300 & WS\tabularnewline
				\hline
				Samsung smartcam & 15000 & 61864 & 567 & SC\tabularnewline
				\hline
				TP-Link smart plug & 15000 & 55372 & 178 & TP\tabularnewline
				\hline 
				Netatmo camera & 15000 & 61859 & 237 & NC\tabularnewline
				\hline 
				Chromecast Ultra & 15000 & 28730 & 252 & CU\tabularnewline
				\hline
				Amazon Echo & 15000 & 35230 & 169 & AE\tabularnewline
				\hline
				Phillips Hue bulb & 15000 & 28730 & 297 & PH\tabularnewline
				\hline
				iHome Smart plug & 15000 & 28730 & 30 & IH\tabularnewline
				\hline
				LiFX bulb & 15000 & 28730 & 150 & LX\tabularnewline
				\hline
			\end{tabular}
		\end{adjustbox}
	\end{minipage} 
	\vspace{-0.4cm}
\end{table*}

In this section we explain our attack scenarios, tool, testbed  and dataset (benign and attack traffic) collected in our lab.

\textbf{Attack Types and Scenarios:} 
We start by designing two attack cases: first, varying rates and location, and next, distributing across various header fields.

\textit{Case 1:} For this case, we consider two types of attacks namely, (a) direct and (b) reflection. Our direct attacks include ARP spoofing, TCP SYN flooding, Fraggle (UDP flooding), and Ping of Death. Reflective attacks include SNMP, SSDP, TCP SYN, and Smurf. IoT devices (\eg WeMo switch and WeMo motion) have limited processing capability and become non-functional when they receive a relatively high rate traffic -- the actual value of a ``high'' rate traffic varies across devices from WeMo motion to Amazon Echo. Also, for reflective attacks, it is important to keep the traffic rate low, ensuring the device remains functional during attack and reflects the attack traffic to the victim -- for example, WeMo switch becomes non-functional under high rate attack traffic, and thus makes the intended attack unsuccessful. Due to these reasons and also to show the ability of our detection method, we use low-rate and high-rate attacks in our experiments.

As depicted in Table~\ref{table:attacsklist}, we launched various types of attacks at different rates, \ie low: 1 packet-per-second (pps), medium: 10 pps, and high: 100 pps, and with diversity of location for both attackers and victims being either from Internet (\ie indicated by ``I'') or local (\ie indicated by ``L'').  In total, we generated 200 of these attacks, each was sustained for a duration of 10 minutes -- note that these attacks were launched over a static number of microflows (not distributed).

\begin{table*}[!t]
	\centering
	\caption{Distributed attacks launched on the IoT devices.}
	\label{table:wormattacks}
	\vspace{-3mm}
	\begin{adjustbox}{max width=0.7\textwidth}	
		\renewcommand{\arraystretch}{1.2}
		\begin{tabular}{|c|c|c|c|c|c|c|c|c|c|c|c|}
			\hline 
			\multirow{2}{*}{Affected Protocol} & \multicolumn{3}{c|}{Packet Rate} & \multicolumn{5}{c|}{Device Label} & \multicolumn{3}{c|}{Affected  Headers}\tabularnewline
			\cline{2-12} 
			& 1 + 0.5 pps & 10 + 1 pps & 100 + 2 pps & WM & WS & SC & NC & AE & sPort & sIP & sPort \& sIP\tabularnewline
			\hline 
			TCP & \cmark& \cmark & \cmark & \cmark & \cmark & \cmark & \cmark &  & \cmark& \cmark& \cmark\tabularnewline
			\hline 
			UDP & \cmark & \cmark & \cmark & \cmark &  & \cmark &  & \cmark & \cmark & \cmark & \cmark\tabularnewline
			\hline 
			ICMP & \cmark & \cmark & \cmark &  & \cmark & \cmark &  &  & N/A & \cmark & N/A\tabularnewline
			\hline 
		\end{tabular}
	\end{adjustbox}
	\vspace{-3mm}
\end{table*}

We designed these specific attacks to analyze how different rates of attack would impact the traffic in various protocols including ARP, TCP, UDP, and ICMP -- note that application layer attacks such as  HTTP, HTTPS, DNS, and SMTP will ultimately affect the behavior of these lower-layer protocols, and hence we do not need to monitor the behavior of application-specific protocols. Our intention was to launch attacks (to the device or Internet servers) without being detected by the specification-based intrusion detector, meaning they conform to the IoT device MUD profiles. 
Furthermore, these attacks were sourced from within the local network as well as from the Internet. For Internet sourced attacks, we enabled port forwarding on the gateway (emulating a malware behavior \cite{Wisec17}). 
For local attacks we employed IP and port spoofing, and for Internet attacks we employed DNS spoofing followed by IP and port spoofing. 



\textit{Case 2: } For this case, we generate distributed attacks by considering three scenarios: (a) varying source port numbers only, while source IP address is fixed, (b) varying IP addresses only, while source port number is fixed, and (c) varying both source ports and IP addresses. As depicted in Table~\ref{table:wormattacks}, these attacks were launched on 5 IoT devices over TCP, UDP and ICMP protocols. In order to control the dynamics of distributed attacks, we sourced them from two machines in three phases, each demonstrating one of the scenarios mentioned above. Attack phases (each lasting for 10 minutes) are performed in parallel on the source machines at three different pairs of rates, namely (1pps, 0.5pps), (10pps, 1pps) and (100pps, 2pps) -- in each pair, values represent the packet rate on their respective machine.
Note that target headers in each phase change by every packet sent --  each packet results in a unique microflow. 

\textbf{Tool:} We developed a modular tool, written in Python, to provide a suite of attacks specific to several real consumer IoT devices that are currently available on the market.
The tool automatically identifies vulnerabilities of a device (SSDP, SNMP, exposed ports, weak encryption or unencrypted communication) by launching various tests against the device on the local network. Once the device’s vulnerabilities are identified, the tool then launches pertinent attacks. During the attacks, the tool generates appropriate annotations including the victim device’s IP address, the attacker host information, start-time, end-time, bitrate, attack protocol, and attack port number.

\textbf{Testbed:} The lower part of Fig.~\ref{fig:testbed} illustrates our testbed that was used to evaluate the performance of our method and system, including a
TPLink gateway with OpenWrt firmware that serves a number of IoT devices, including a WeMo switch, a WeMo motion sensor, a Samsung smart-camera, a TP-Link smart plug, a Netatmo camera, a Chromecast Ultra, an Amazon Echo, a LiFX bulb, a Phillips Hue bulb and an iHome Smart plug. Two attackers were included, locally (inside LAN) and remotely (on the Internet) with two victims, both attackers being able to attack both victims.


We connected a 1 TB external hard disk to the gateway to store packet trace (\ie {\myverb{pcap}} files) of all network traffic (\ie locally and remotely) using the {\myverb{tcpdump}} tool. 
Packet traces of benign and attack traffic from the testbed were collected for a period of 16 days. Given the known attackers and victims, it was easy to annotate the attack traffic in the dataset, as shown in Table~\ref{table:summaryexperiment}. Interestingly, the dataset revealed the presence of other attacks launched from the Internet (\ie wild attacks) when port forwarding was enabled. 
In order to capture the benign behavior of IoT devices in our testbed, we installed a touch replay tool on a Samsung galaxy tab recording all possible user interactions (\eg turning on/off lightbulb, or streaming video from camera) with individual IoTs -- each device has a limited number of functions available. We then replayed recorded interactions (spread randomly over hours of day) emulating a real personal activity. For Amazon Echo specifically, we used a simple text-to-speech program that randomly picks a statement from a pre-configured list (\eg ``\textit{Alexa! How is the weather today?}", ``\textit{Alexa! Play a music named X}", etc). 

\textbf{Dataset:} 
We developed two datasets namely, raw packet traces and derived flow counters.
We release \cite{attackdata} our datasets (spanning one month period of benign and attack traffic relating to ten IoT devices and annotation of those attacks).
The released datasets contain 35 pcap files and each file corresponds to a trace collected over a day. Note that there are 17 other IoT devices (\eg TPLink camera, DLink camera) in our testbed that are not studied or experimented in this work but the benign data of these devices are included in our traces -- we only focused on selected devices with more complex behavior. In addition, we release two annotation files comprising: (a) start-time, end-time, flows that are influenced during the attack, attack type, bitrate of attack; and (b) pcap file number, attacker, and victim IP address.
Our derived dataset contains counters of flows (computed over a minute) for 10 IoT devices listed in Table~\ref{table:summaryexperiment}.
The second column shows the number of training instances (\ie count of packets and bytes per flow rule per minute) for each device. The training instances only contain benign traffic. For the testing phase, 17420 instances were collected for each device, containing both benign and attack traffic. Of these testing instances, the number of attack instances is shown in the fourth column (in Table~\ref{table:summaryexperiment}) for the corresponding IoT device under attack.


\vspace{-0.3cm}
\section{Prototype and Evaluation}\label{sec:evaluation} 
\begin{figure}[!t]
	\centering
	\includegraphics[width=0.87\linewidth]{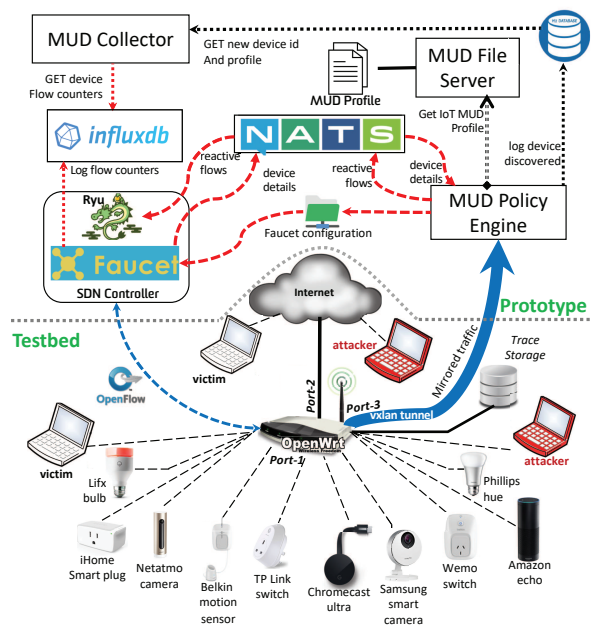}
	\vspace{-4mm}	
	\caption{Our system prototype and testbed.}
	\label{fig:testbed}
	\vspace{-6mm}
\end{figure}

We prototyped our scheme in a small testbed, depicted in Fig.~\ref{fig:testbed}. The objectives of this experimental setup are to demonstrate the feasibility of our scheme with real equipment and traffic, and to evaluate the efficacy of anomaly detection and attack identification. 

\vspace{-0.4cm}
\subsection{Prototype Implementation}\label{sec:implementation} 
For our system. we developed an application on top of the open-source  Ryu along with the Faucet/Gauge \cite{faucet} SDN controllers, the MUD policy engine, the MUD collector, and implemented the NATS messaging system and used the InfluxDB and H2 databases. Each of these components operates on a separate docker container over an Ubuntu 16.04 server. In addition, the MUD file server is a repository of MUD profiles (obtained from \cite{IoTSnP18-mudgee, hamza2020verifying}) that runs as an HTTP server on a separate VM in our University cloud \cite{unswmudrepo}. We release our prototype as open-source \cite{mudie}. Technical details of the system components are provided in our previous work \cite{MUDlearn}.

\subsection{Feature Analysis} 
We now evaluate the importance of features in performance of our anomaly detection considering both feature types.

\subsubsection{Volumetric features} 
As explained in \S\ref{sec:featureExtract}, we collect flow statistics every minute and construct features using these statistics in three possible scenarios: (a) \textbf{feature-set-1 (FS1)}: only total count over sliding windows (\eg for window size of 2 min, features are 1-min and 2-min total count of flow bytes and packets), (b) \textbf{feature-set-2 (FS2)}: total count for the last one minute, and mean and standard deviation over the window (\eg for a 2-min window size, features are 1-min total count, 2-min mean and 2-min standard deviation of byte/packet counts), and (c) \textbf{feature-set-3 (FS3)}: a combination of FS1 and FS2. Note that for a 1-min sliding window, all FS1, FS2, and FS3 correspond to the same set of features.
In Fig.~\ref{fig:featureanalysis}, we plot F-score (\ie a measure of accuracy in binary classification), True Positive rate (TPR), False Positive rate (FPR) for each of these feature sets when sliding window varies from 1 to 8 minutes. This figure also illustrates the performance of two anomaly detection techniques: (i) only boundary detection (BD), and (ii) BD combined with state machine.

\newlength{\figurewidthA}
\newlength{\figureheigthA}
\setlength{\figurewidthA}{0.3\textwidth}
\setlength{\figureheigthA}{0.16\textwidth}
\begin{figure*}[t!]
	\centering
	\subfigure[F-Score.] 
	{
		\includegraphics[width=\figurewidthA,height=\figureheigthA]{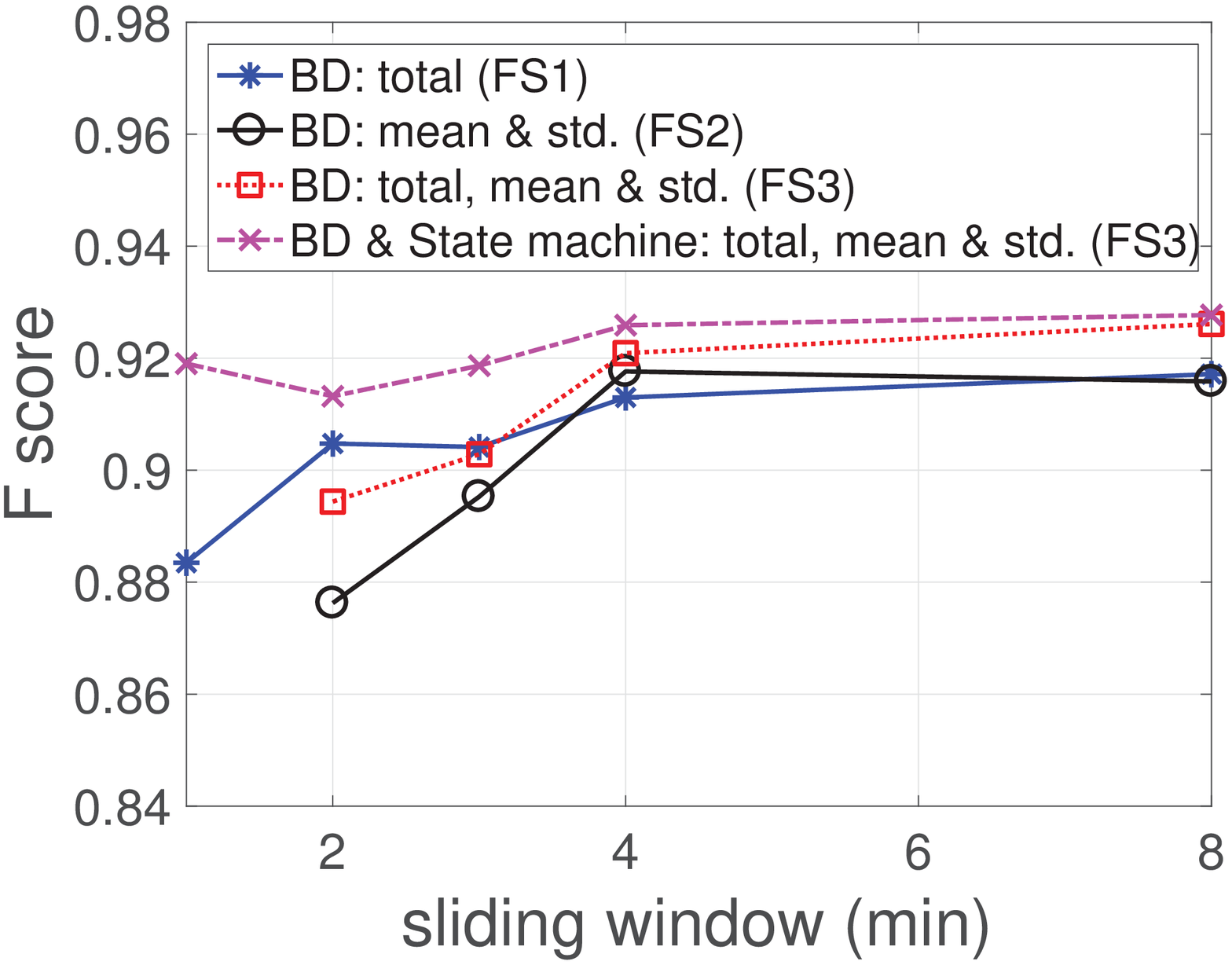}
		\label{fig:f1scoreresult}
	}
	\subfigure[True Positive rate.] 
	{
		\includegraphics[width=\figurewidthA,height=\figureheigthA]{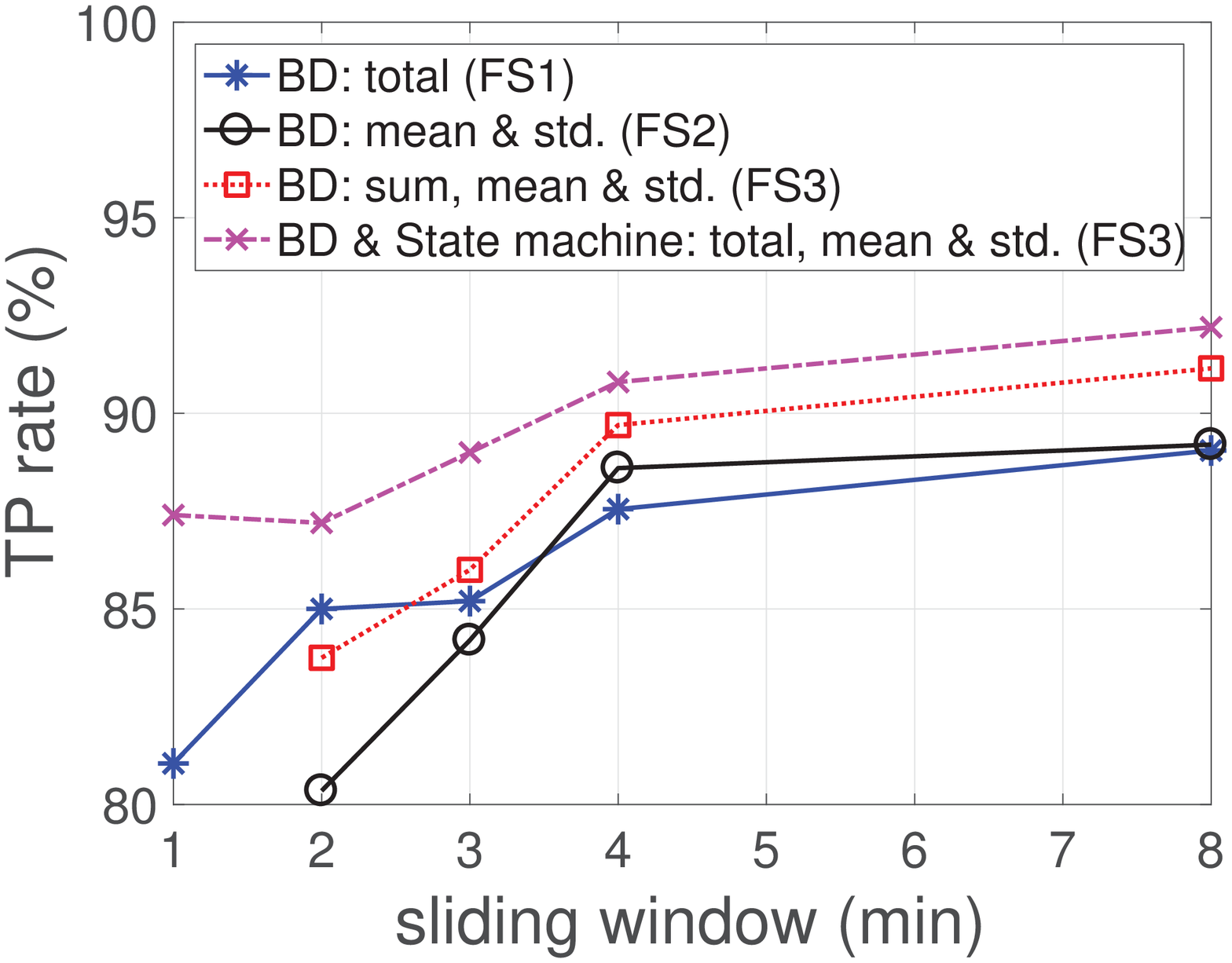}
		\label{fig:tprresult}
	}
	\subfigure[False Positive rate.] 
	{
		\includegraphics[width=\figurewidthA,height=\figureheigthA]{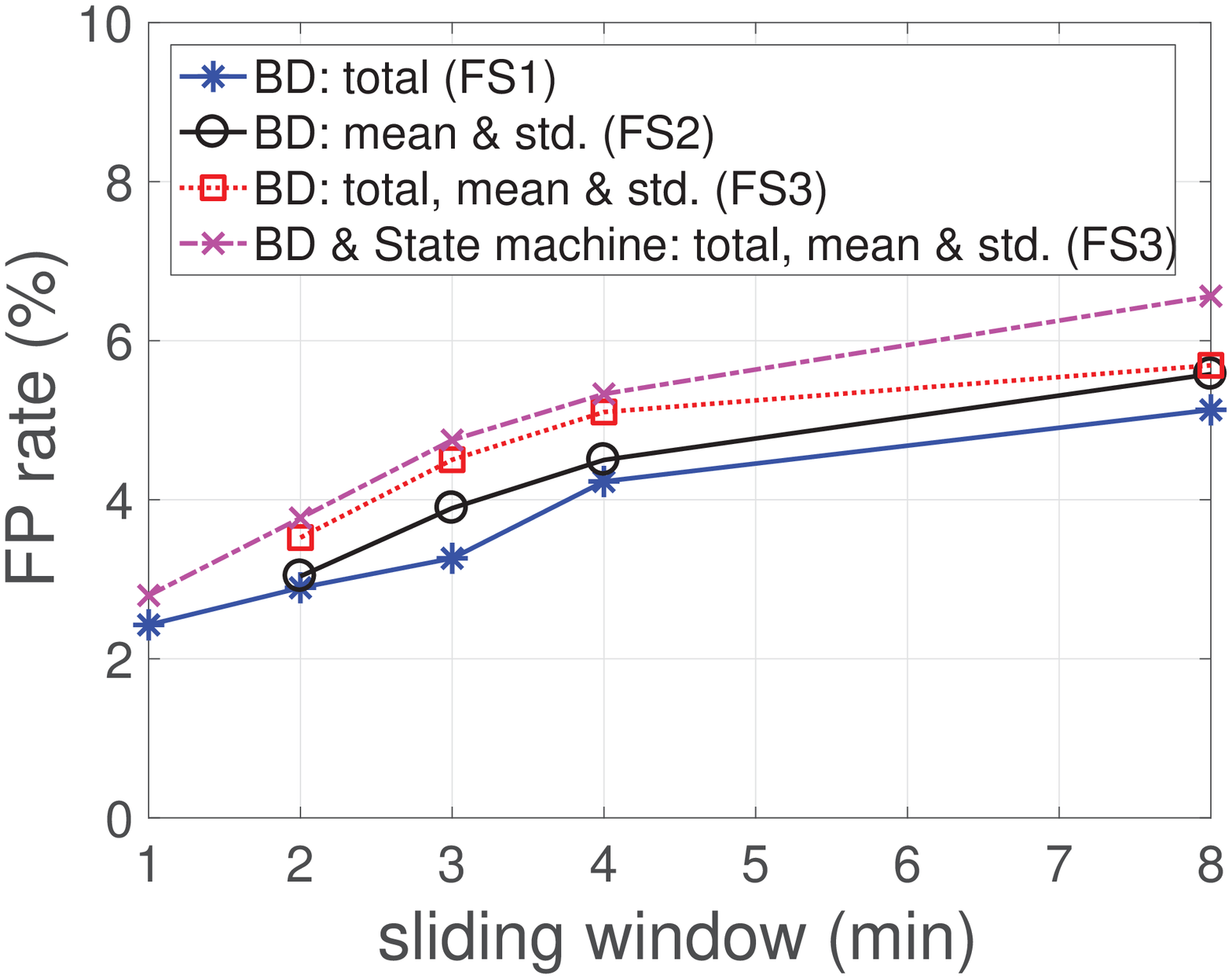}
		\label{fig:fprresult}
	}
	\vspace{-3mm}
	\caption{Impact of various features on performance of volumetric anomaly detection (with boundary detection).}
	\label{fig:featureanalysis}
	\vspace{-4mm}
\end{figure*}

\textbf{Impact of window size:} According to Fig.~\ref{fig:f1scoreresult}, for BD only, as the window size increases the performance is improved steadily, with FS3 outperforming FS1 and FS2. Note FS1 performs better in smaller windows sizes, since mean/ standard deviation would not give extra information for small number of data points. Looking into Figures~\ref{fig:tprresult}~and~\ref{fig:fprresult}, we observe that a larger window size results in a high rate of true positives and false positives. In order to detect low rate attacks, we need to choose a larger window size -- large windows impose computing costs and demand higher memory footprint figures.  

\textbf{Impact on detection:} Use of just state machine (without boundary detection) results in $54.55$\% of TPR and $0.69$\% of FPR with 1 minute window for FS1 -- a significant portion of attacks are missed. Applying just boundary detection, instead, gives a high TPR $81.05$\%. Combining both boundary detection and state machine gives $87.40$\% of TPR (with 1 min window) -- 5 attacks (from a total of 200) that are low rate get missed. The TPR can be further improved to $90.80$\% by incorporating a richer feature set FS3 and increasing the sliding window to 4 minutes (all low-rate attacks get detected) -- this is gained at the cost of higher FPR in Fig.~\ref{fig:fprresult}. We note that this TPR can be achieved by employing just boundary detection with FS3 and window size of 4 min (as shown by red lines with square markers in Fig.~\ref{fig:tprresult}). This shows that traffic characteristics captured by the state machine can be captured by the boundary detection but over a larger window.

\textbf{Summary:} When low-rate attacks are not of interest to the network operator then smaller window sizes with FS1 using both boundary detection and state machine are recommended (lower cost and better FPR). If the operator wants to detect all possible attacks (both low rates and high rates), a large window size with FS3 (using only boundary detection) would be an efficient approach -- use of a larger window comes with the cost of maintaining states for computing features and results in a slightly higher false positive rate.  In what follows, we will employ the latter approach to detect all attacks and operate over the sliding window of 4 minutes -- going beyond 4 minutes does not significantly affect TPR or FPR, but it requires double the amount of states. Note that this does not impact on the responsiveness of our detection method -- it still responds every one minute. 

\begin{table}
	\centering
	\caption{Performance of dispersion detector as a function of $\break$epoch duration, across all devices.}
	\label{table:entropyperformance}
	\vspace{-3mm}
	\begin{adjustbox}{max width=0.48\textwidth}	
		\renewcommand{\arraystretch}{1.2}			
		\begin{tabular}{|c|c|c|c|}
			\hline 
			Epoch duration (sec) & Accuracy (\%) & TPR (\%) & FPR (\%)\tabularnewline
			\hline 
			\hline 
			1 & 97.1 & 79.7 & 2.3\tabularnewline
			\hline 
			5 & 98.3 & 98.7 & 1.7\tabularnewline
			\hline 
			10 & 93.5 & 98.9 & 6.5\tabularnewline
			\hline 
		\end{tabular}
	\end{adjustbox}	
	\vspace{-6mm}
\end{table}

\subsubsection{Dispersion features} In distributed attacks, it is essential to detect the attack quickly to avoid TCAM exhaustion. Therefore, we considered epochs including 1 sec, 5 sec, and 10 sec for monitoring the entropy of dynamic headers. Results are shown in Table~\ref{table:entropyperformance}. It can be seen that increasing the epoch duration would increase both TPR and FPR. We note that longer epochs reduce the cost of computing entropy, but it may lead to TCAM exhaustion due to insertion of reactive microflows. Therefore, in real practices, network operators need to configure an appropriate epoch, considering their switch TCAM capacity as well as their compute resources available for the packet inspection engine. 


\vspace{-3mm}
\subsection{Attack Detection}\label{attackdetx}

\begin{table*}[!t]
	\centering
	\caption{Performance of our anomaly detectors at stage-1 and stage-2.}
	\label{table:performance}
	\vspace{-3mm}
	\begin{adjustbox}{max width=0.7\textwidth}	
		\renewcommand{\arraystretch}{1.2}			
		\begin{tabular}{|l|c|c|c||c|c|c|c|c|c|c|c|c|c|c|c|}
			\hline 
			\multirow{2}{*}{Anomaly Detectors} & \multicolumn{3}{c||}{All devices} & \multicolumn{2}{c|}{WM} & \multicolumn{2}{c|}{TP} & \multicolumn{2}{c|}{SC} & \multicolumn{2}{c|}{NC} & \multicolumn{2}{c|}{CU} & \multicolumn{2}{c|}{AE}\tabularnewline
			\cline{2-16} 
			& \rotatebox{90}{Accuracy (\%)\hspace{0.3em}}  & \rotatebox{90}{TPR (\%)} & \rotatebox{90}{FPR (\%)} & \rotatebox{90}{TPR (\%)} & \rotatebox{90}{FPR (\%)} & \rotatebox{90}{TPR (\%)} & \rotatebox{90}{FPR (\%)} & \rotatebox{90}{TPR (\%)} & \rotatebox{90}{FPR (\%)} & \rotatebox{90}{TPR (\%)} & \rotatebox{90}{FPR (\%)} & \rotatebox{90}{TPR (\%)} & \rotatebox{90}{FPR (\%)} & \rotatebox{90}{TPR (\%)} & \rotatebox{90}{FPR (\%)}\tabularnewline
			\hline 
			Stage-1 and Stage-2 combined & \textbf{94.9} & \textbf{89.7} & \textbf{5.1} & 95.7 & 5 & 95.7 & 2.3 & 93.8 & 3.1 & 80.3 & 4.2 & 79.8 & 19.7 & 83.5 & 3.8\tabularnewline
			\hline 
			Stage-1 \& -2 combined (2-min filtering) & 97.5 & \textbf{72.3} & \textbf{2.4 }& 77.7 & 2.4 & 76 & 1 & 75.6 & 0.9 & 71.4 & 2.1 & 63.1 & \textbf{13.2} & 67.1 & \textbf{0.6}\tabularnewline
			\hline 
			Only Stage-2 detector  & 89.1 & 92 & 10.9 & 96 & 11.3 & 96.2 & 3.4 & 94.1 & 6.5 & 83.7 & 6.1 & 86.1 & 52.3 & 91.1 & 16.3\tabularnewline
			\hline 
			Only Stage-1 detector  & 85.7 & 93.7 & 14.4 & 97 & 14.8 & 96.2 & 3.9 & 98.3 & \textbf{30.5} & 88.4 & 8.7 & 88.9 & 27.7 & 93.7 & 23.5\tabularnewline
			\hline 
			Only Local detector & 90.6 & 91.5 & 9.4 & 87.1 & 8.6 & 96 & 2.2 & 96.2 & 28.4 & 94.3 & 1.5 & \textbf{80.7} & 7.1 & 93.3 & 15.9\tabularnewline
			\hline 
			Only Internet detector  & 93.9 & 88.5 & 6.1 & 95 & 7.2 & 93.2 & 2.1 & 94 & 3.3 & 75 & 8 & 84.8 & 23.2 & \textbf{68.4} & 9.2\tabularnewline
			\hline 
		\end{tabular}
	\end{adjustbox}
	\vspace{-3mm}
\end{table*}

\begin{table*}[!t]
	\centering
	\caption{Detected anomalous MUD flow in IoT attacks. (\textit{L:local, d:device, I:Internet})}
	\label{table:attackdetectors}
	\vspace{-3mm}
	\begin{adjustbox}{max width=0.70\textwidth}	
		\renewcommand{\arraystretch}{1.2}			
		\begin{tabular}{|l|c|c|c|c||c|c|c||c|c|c||c|c|c||c|c|c||c|c|c|}
			\hline 
			\multirow{2}{*}{Attack Type} & \multirow{2}{*}{Attack Scenario} & \multicolumn{3}{c||}{All device} & \multicolumn{3}{c||}{WM} & \multicolumn{3}{c||}{TP} & \multicolumn{3}{c||}{SC} & \multicolumn{3}{c||}{NC} & \multicolumn{3}{c|}{CU}  \tabularnewline
			\cline{3-20} 
			&  & \rotatebox{90}{Launched } & \rotatebox{90}{Detected} & \rotatebox{90}{Identified} & \rotatebox{90}{Launched} & \rotatebox{90}{Detected} & \rotatebox{90}{Identified} & \rotatebox{90}{Launched} & \rotatebox{90}{Detected} & \rotatebox{90}{Identified} & \rotatebox{90}{Launched} & \rotatebox{90}{Detected} & \rotatebox{90}{Identified} & \rotatebox{90}{Launched} & \rotatebox{90}{Detected} & \rotatebox{90}{Identified} & \rotatebox{90}{Launched} & \rotatebox{90}{Detected} & \rotatebox{90}{Identified}\tabularnewline
			\hline 
			TCP SYN reflection & L$\rightarrow$d$\rightarrow$L & 208 & 198 & 178 & 30 & 30 & {\color{blue}\textbf{21}} & 29 & 28 & 28 & 30 & 29 & 20 & 30 & 29 & 28 & 30 & 29 & 28\tabularnewline
			\hline 
			TCP SYN reflection & I$\rightarrow$d$\rightarrow$I & 221 & 186 & 177 & 30 & 28 & 28 & 30 & 29 & 28 & 30 & 28 & 27 & 30 & {\color{red}\textbf{20}}  & 17 & 41 & 24 & 21\tabularnewline
			\hline 
			SSDP reflection & L$\rightarrow$d$\rightarrow$L & 90 & 63 & 18 & 30 & 26 & 18 &  &  &  &  &  &  &  &  &  & 30 & {\color{red}\textbf{12}}  & 0\tabularnewline
			\hline 
			SSDP reflection & I$\rightarrow$d$\rightarrow$I & 91 & 89 & 85 & 30 & 29 & 29 &  &  &  &  &  &  &  &  &  & 31 & 31 & 27\tabularnewline
			\hline 
			SSDP reflection & L$\rightarrow$d$\rightarrow$I & 89 & 85 & 76 & 30 & 29 & 29 &  &  &  &  &  &  &  &  &  & 30 & 28 & 19\tabularnewline
			\hline 
			SNMP reflection & L$\rightarrow$d$\rightarrow$L & 27 & 22 & 0 &  &  &  &  &  &  & 27 & 22 & 0 &  &  &  &  &  & \tabularnewline
			\hline 
			SNMP reflection & I$\rightarrow$d$\rightarrow$I & 30 & 28 & 28 &  &  &  &  &  &  & 30 & 28 & 28 &  &  &  &  &  & \tabularnewline
			\hline 
			SNMP reflection & L$\rightarrow$d$\rightarrow$I & 30 & 29 & 29 &  &  &  &  &  &  & 30 & 29 & 29 &  &  &  &  &  & \tabularnewline
			\hline 
			Smurf & L$\rightarrow$d$\rightarrow$L & 120 & 102 & 97 &  &  &   &  30 & 27  & 27 & 30 & 28 & 27 &  &  &  &  &  & \tabularnewline
			\hline 
			Fraggle & L$\rightarrow$d & 120 & 110 & 109 & 30 & 29 & 29 &  &  &  & 30 & 28 & 28 &  &  &  &  &  & \tabularnewline
			\hline 
			Fraggle & I$\rightarrow$d & 108 & 85 & 79 &  &  &  &  &  &  & 30 & 29 & 29 &  &  &  &  &  & \tabularnewline
			\hline 
			TCP SYN & L$\rightarrow$d & 210 & 203 & 184 & 30 & 30 & 21 & 30 & 30 & 30 & 30 & 29 & 20 & 30 & 28 & 28 & 30 & 30 & 29\tabularnewline
			\hline 
			TCP SYN & I$\rightarrow$d & 179 & 145 & 139 & 30 & 29 & 28 & 29 & 27 & 26 & 30 & 28 & 27 & 30 & {\color{red}\textbf{16}}  & 14 & 30 & {\color{red}\textbf{18}}  & 16\tabularnewline
			\hline 
			Arp Spoof & L$\rightarrow$d & 297 & 282 & 276 & 30 & 29 & 29 & 30 & 28 & 28 & 30 & 28 & 28 & 27 & 25 & 24 & 30 & 29 & 28\tabularnewline
			\hline 
			Ping of death & L$\rightarrow$d & 180 & 167 & 158 & 30 & 28 & 28 & 30 & 30 & 30 & 30 & 29 & 29 &  &  &  &  &  & \tabularnewline
			\hline 
		\end{tabular}
	\end{adjustbox}
	\vspace{-4mm}
\end{table*}

It is paramount for our scheme to detect anomalies at stage-1 and stage-2, for an improved inferencing at stage-3 in determining the nature of the attack and identifying the anomalous microflows. In this subsection, we analyze the performance of stage-1 and stage-2 models, and \S\ref{sec:attack5tuple} will focus on the performance of stage-3 workers. Note that our inferencing at stage-3 would be relatively expensive (packet inspection and TCAM consumption), and therefore, low FPRs in both stage-1 and stage-2 are highly desirable. The results for aggregate of all devices and individual IoT device types are reported in Table~\ref{table:performance}.


\textbf{Accuracy and False-Positive Rate:}
Focusing on aggregate of all devices, it is seen that the combination of the two stages yields the highest accuracy of $94.9$\% (\ie the percentage of correctly classified benign and attack instances). We are able to detect $89.7$\% of all attacks (TPR: true positive rate) across all IoT devices, when the initial two-stage anomaly detection is employed. As we expect, in this situation the lowest false positive rate $5.1$\% is achieved. 
Even this FPR may not be very attractive for real network settings with a large number of connected devices. To reduce FPR, one can employ a \textbf{\textit{time-based filtering}}. We use a simple threshold for raising alarms if the anomaly detection is triggered continuously for more than ``\textit{t}'' minutes. As shown in the second row of Table~\ref{table:performance}, having a 2-minute filter reduces the FPR to $2.4$\%. However, the TPR is also reduced to $72.3$\% -- because attacks were not detected during their first two minutes due to time-based alarm filtering. Increasing this time threshold would enhance the FPR but it is detrimental to detection responsiveness.



Unsurprisingly, when workers of only stage-1 or stage-2 are used, the overall accuracy drops. We found that stage-2 workers perform slightly better than stage-1 workers, however in terms of functionality both have separate purposes. Stage-1 inferencing deals with coarse-grained device-level activity whereas stage-2 deals with fine-grained  service-level activity. Thus, a combined inferencing provides better accuracy and and fewer false-positive detections.

Considering per-device performance of anomaly detection, the bottom two rows in Table~\ref{table:performance} show performance when local and Internet attacks are separately considered. For the local detector, the lowest true positive rate (\ie TPR $80.7$\%) is achieved for the Chromecast Ultra (\ie device label ``CU''). We found that some of our reflection attacks originated from the local attacker to an external victim (\ie L$\rightarrow$d$\rightarrow$I) are missed by this worker, meaning that local traffic features are not impacted sufficiently to raise an attack alarm. However, these reflection attack instances are detected by the Internet worker. 
Similarly, we observe that the Internet detector for Amazon Echo (\ie device label ``AE'') suffers from a fairly low TPR of $68.4$\% -- only a few instances of 1 pps and 10 pps TCP SYN attacks from Internet were missed, but ultimately both of these attacks were detected with some delays. Overall, our models were able to successfully detect all types (high-, med-, low-rate) of attacks.

Another interesting observation is that for the stage-1 (device-level traffic) worker of Samsung camera (\ie device ``SC''), the false positive rate is very high ($30.5$\%); however, when combined with the stage-2 (service-level traffic) inferencing, the false-positives drop to $3.1\%$. This shows that the coarse-grained behavior (\ie aggregate of flows) of this device was not fully learned by the training dataset, but service-level behavior was well captured and learned.


\textbf{Detecting Various Attack Types:}
In Table~\ref{table:attackdetectors} we show the number of detected attack instances for each IoT device per attack type -- each instance represents a one minute period of traffic. For example in the first row, we launched 30 instances of TCP SYN remote reflection attacks (\ie L$\rightarrow$d$\rightarrow$L) to device label ``A'', and the anomaly detection machine was able to detect all of these 30 attacks just one minute after their commencement
The results shown in Table~\ref{table:attackdetectors} highlight the fact that our method is able to detect volumetric attacks of all types during their lifetime (\ie 10 minutes or more).

\begin{table*}[!t]
	\caption{Identified anomalous MUD flow for TP-Link smart plug under attack.}
	\centering
	\label{table:tplinkattackflows}
	\vspace{-0.3cm}
	\begin{adjustbox}{max width=0.68\textwidth}	
		\renewcommand{\arraystretch}{1.2}			
		\begin{tabular}{|l|c|c|c|c|c|c|c|c|c|c|c|c|}
			\hline 
			\multirow{2}{*}{\textbf{Attack type}} & \multirow{2}{*}{\textbf{Attack scenario}} & \multirow{2}{*}{\textbf{Launched}} & \multirow{2}{*}{\textbf{Detected}} & \multirow{2}{*}{\textbf{Malicious flow}} & \multicolumn{8}{c|}{\textbf{Identified flows}}\tabularnewline
			\cline{6-13} 
			&  &  &  &  & a & b & d & e & f & h & i & j\tabularnewline
			\hline 
			TCP SYN reflection & L$\rightarrow$d$\rightarrow$L & 29 & 28 & i & 0 & 0  & 0 & 0 & 0 & {\color{red}28} & 28 & 0\tabularnewline
			\hline 
			TCP SYN reflection & I$\rightarrow$d$\rightarrow$I & 30 & 29 & b & 0 & 28  & 0 & 0 & 3 & 3 & 0 & 3\tabularnewline
			\hline 
			Smurf & L$\rightarrow$d$\rightarrow$L & 30 & 27 & j & 0 & 0  & 0 & 0 & 0 & {\color{red}27} & 0 & 27\tabularnewline
			\hline 
			TCP SYN & L$\rightarrow$d & 30 & 30 & i & 0 & 0  & 0 & 0 & 0 & {\color{red}30} & 30 & 0\tabularnewline
			\hline 
			TCP SYN & I$\rightarrow$d & 29 & 27 & b & 0 & 26 & 0 & 0 & 0 & 0 & 0 & 0\tabularnewline
			\hline
			Ping of death & L$\rightarrow$d & 30 & 30 & e & 0 & 0  & 0 & 30 & 0 & {\color{red}26} & 0 & 0\tabularnewline
			\hline
			ARP spoof & L$\rightarrow$d & 30 & 28 & h & 0 & 0  & 0 & 0 & {\color{blue}15} & 28 & 0 & {\color{blue}12} \tabularnewline
			\hline 
		\end{tabular}
		\vspace{-7mm}
	\end{adjustbox}
	\vspace{-6mm} 
\end{table*}

We note that our scheme may miss certain types of reflection attacks within the local network (\ie L$\rightarrow$d$\rightarrow$L), namely those that are broadcast with the source address spoofed as a local victim. For this specific type, the original attack traffic does not match on any device-specific flow rule (\eg SNMP reflection attack for Samsung camera and SSDP reflection for Chromecast Ultra) because our system only captures incoming traffic for the local network (as explained in \S\ref{sec:design}).
But the reflected traffic may contribute to one of the device flows. For example, it was found that the local SSDP reflection attack on the WeMo motion detector device was detected because the reflected packets happened to match one flow of the WeMo motion. Note that if the victim is an IoT device, even the local broadcast reflection attack will be detected.


There is also a low detection, shown by red text in Table~\ref{table:attackdetectors}, for SSDP reflection (L$\rightarrow$d$\rightarrow$L) and TCP SYN (I$\rightarrow$d) in Chromecast Ultra(\ie device ``CU'')
, and TCP SYN reflection (I$\rightarrow$d$\rightarrow$I) and TCP SYN (L$\rightarrow$d) in Netatmo camera (\ie device ``NC''). 
We emphasize that these undetected attack instances were for low-rate traffic (each one minute) from a 10-minute duration of an attack. This indicates that low-rate attacks are difficult to detect, but the detection process is able to detect them if their duration is long, which is typically the intention of the attacker. In other words, all of the attacks, were ultimately detected (\ie on average after 1.92 minutes from the attack commencement).  
%

\begin{figure}[t!]
	\centering
	\subfigure[perofrmance vs. training.] 
	{
		\includegraphics[width=0.2255\textwidth,height=0.14\textwidth]{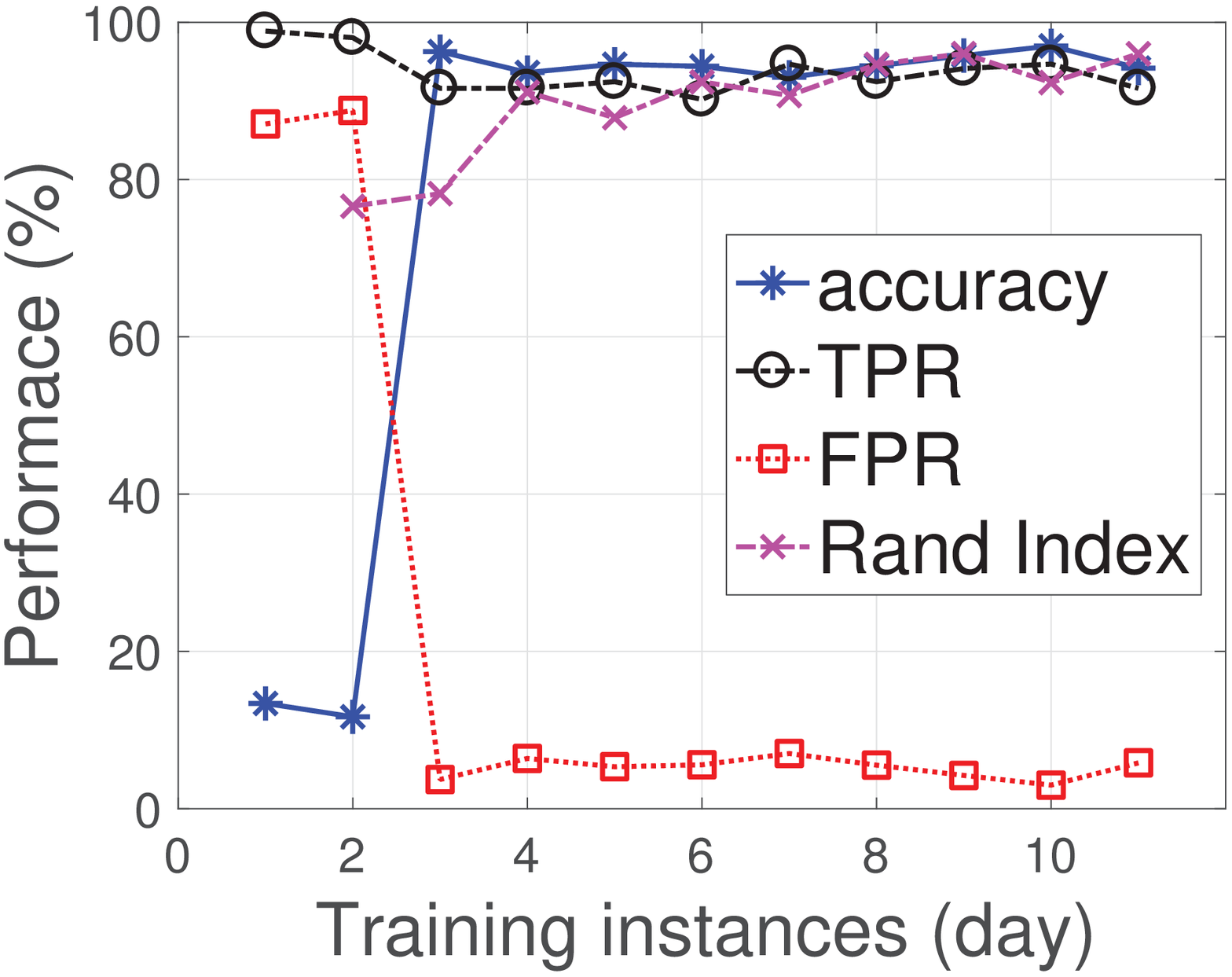}
		\label{fig:samsungperf}
	}
	\subfigure[ARP profile.] 
	{
		\includegraphics[width=0.2255\textwidth,height=0.14\textwidth]{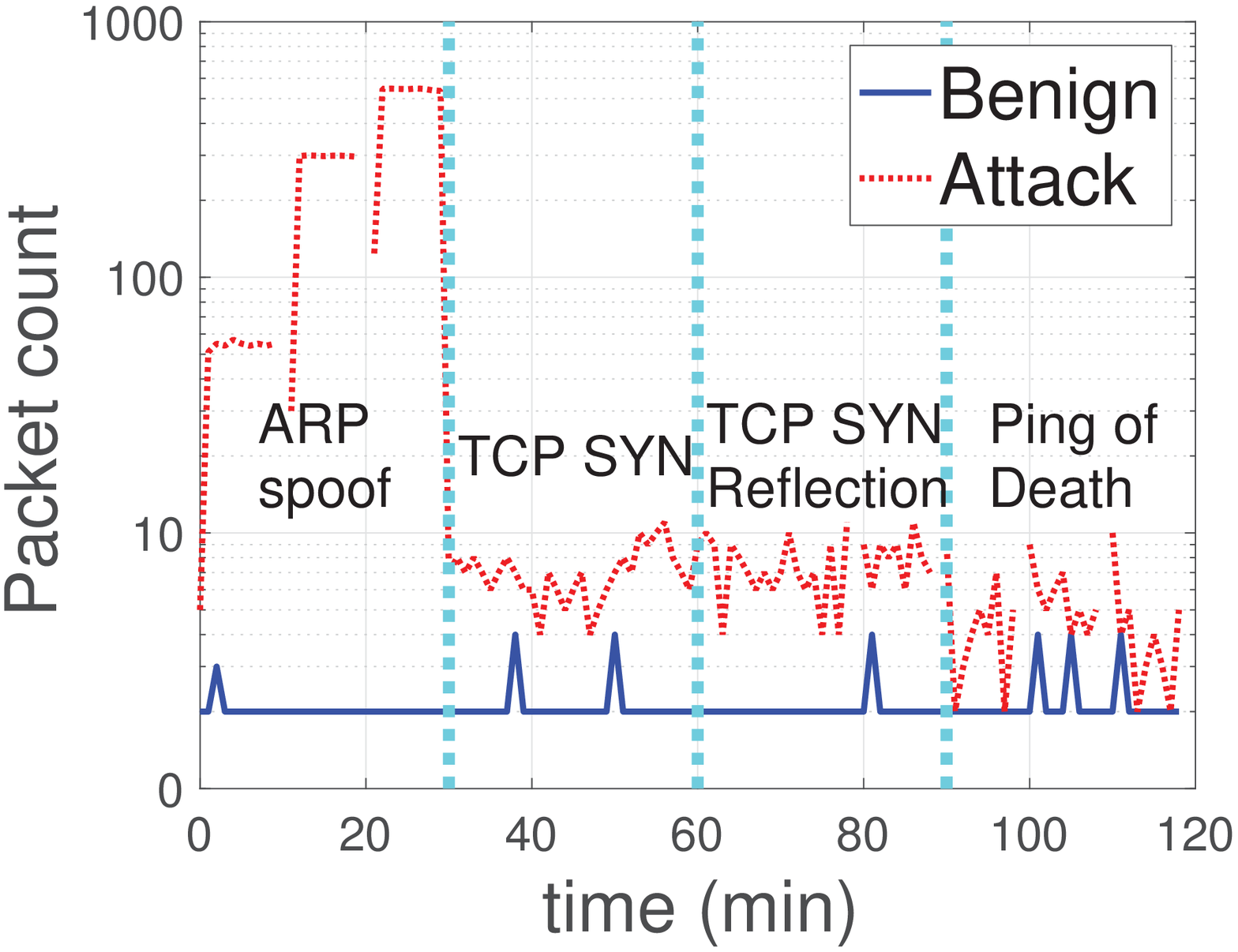}
		\label{fig:tplinkarp}
	}
	\vspace{-4mm}
	\caption{(a) Performance of anomaly detection for Samsung camera, (b) ARP traffic profile for TP-Link smart plug (benign versus local attack).}
	\label{fig:results}
	\vspace{-6mm}
\end{figure}

\textbf{Impact of Training on Performance:} The accuracy of our attack detection highly depends on the benign states that are learned during the training phase. We see in Fig.~\ref{fig:samsungperf} that the overall accuracy for Samsung camera is less than $50$\% when the model is trained by only 2-days of training data, and steeply rises to $96.08$\% when models are trained with one additional day of training instances that include new benign (expected) states. 
In contrast, the TPR rate is consistently high (\ie above 80\%) because all attack instances (including low rate ones) deviate from even limited numbers of trained states. Therefore, it is important to capture all benign states (\ie normal traffic) of each device during the training phase. It is challenging to determine the minimum amount of training data (in terms of the number of days) for building a reasonably well-trained model.


We use ``Rand index'' \cite{hubert1985comparing} (a measure of the similarity between two data clusters) to identify the minimum amount of training dataset (in terms of number of days) for building a well-trained model. 
It is shown by dashed pink lines with cross markers in Fig.~\ref{fig:samsungperf}. A consistently high Rand index indicates that the training data is sufficient. We can see that 4 days of training instances would result $91$\% of Rand index and will relatively persist with more instances trained. 
\subsection{Identifying Attack Service Flow}
We now look at the performance of our scheme in identifying attacks at service flow level.
In Table~\ref{table:attackdetectors}, the ``Identified'' column under each device shows the number of attack instances in which the contributing flow was correctly identified. It can be seen that for TCP SYN local reflections (L$\rightarrow$d$\rightarrow$L) in the first row, of which all 30 instances were detected, but only in 21 of those was the attack flow identified correctly.
In the remaining 9 instances, only the ARP flow was flagged –- however, the attack was not launched over ARP.  It was found that the ARP anomaly worker is sensitive to (\ie raises alarms for) most local attacks, highlighted by bold text  under column ``Identified'' under the WeMo motion (``WM'') heading in  Table~\ref{table:attackdetectors}, while the actual contributing flow was not identified for some instances. In the case of the Fraggle and Ping-of-death attacks, the corresponding attack flow was correctly flagged by the worker, although the ARP worker once again incorrectly flagged the ARP flow.
The packets that match anomalous MUD flows get mirrored to our packet inspection engine, to further investigating the microflows that cause the anomaly (\S\ref{attackflowiden}). 



\textbf{Correctness of Flow Alarms:}
The performance of individual flow classifiers/workers (\ie in the stage-2 anomaly detectors) can be seen in Table~\ref{table:tplinkattackflows}, which lists detected attacks and corresponding flows identified for the TP-Link smart plug device. 
The ``Malicious Flow'' column shows the flow (from Table~\ref{table:devrules}) used in the attack. 
For TCP SYN reflection (L$\rightarrow$d$\rightarrow$L), we used TCP port 9999 (flow \textit{i}). It is seen that 28 out of 29 attack instances were correctly detected and all true alarms flagged the correct flow \textit{i}, though ARP flow (\ie flow \textit{h}) is also flagged in 28 alarms. Such a high rate of incorrect ARP alarms is seen for local TCP SYN,  Smurf, Ping-of-Death, highlighted in red text. In order to better understand the reason for ARP alarms, we plot in  Fig.~\ref{fig:tplinkarp} the time profiles of ARP flow in benign (solid blue line) versus in local attack (dashed red line) traffic from the training dataset. 
It is clearly seen that that the ARP profile during an attack clearly deviates from its normal profile, even for attacks that are not directly related to this flow. It is noted that, during Internet attacks, the device ARP profile is not impacted significantly to raise alarms.

\begin{table*}[!t]
	\centering
	\caption{Performance of stage-3 anomaly identification.}
	\label{table:stage3performance}
	\vspace{-3mm}
	\begin{adjustbox}{max width=0.9\textwidth}	
		\renewcommand{\arraystretch}{1.2}			
		\begin{tabular}{|l|c|c|c||c|c|c|c|c|c|c|c|c|c|c|c|}
			\hline 
			\multirow{2}{*}{} & \multicolumn{3}{c|}{All devices} & \multicolumn{2}{c|}{WM} & \multicolumn{2}{c|}{TP} & \multicolumn{2}{c|}{SC} & \multicolumn{2}{c|}{NC} & \multicolumn{2}{c|}{CU} & \multicolumn{2}{c|}{AE}\tabularnewline
			\cline{2-16} 
			& Accuracy & TPR & FPR & TPR & FPR & TPR & FPR & TPR & FPR & TPR & FPR & TPR & FPR & TPR & FPR\tabularnewline
			\hline 
			(i) (stage-1 + stage-2) + stage-3 [dynamically] & 98.7 & 85.0 & 1.2 & 92.0 & 2.6 & 92.8 & 0.2 & 86.2 & 0.6 & 78.9 & 0.7 & 68.2 & 5.0 & 82.3 & 1.0\tabularnewline
			\hline 
			(ii) Only stage-3 [permanently] & 91.3 & 91.0 & 8.6 & 95.3 & 32.3 & 94.2 & 0.5 & 88.5 & 4.3 & \textbf{95.9} & 4.2 & 81.3 & 20.0 & 89.8 & 14.2\tabularnewline
			\hline 
			
		\end{tabular}
	\end{adjustbox}
	\vspace{-5mm}
\end{table*}

Another interesting observation is when the ARP spoofing attack is launched (the last row in Table~\ref{table:tplinkattackflows}). We see 15 and 12 alarms, respectively for DNS (flow \textit{f}) and local ICMP (flow \textit{j}). 
The ARP spoof causes all victim traffic to be redirected to the attacker (instead of the expected gateway). Since the TP-Link smart plug device was communicating ICMP and DNS packets during the ARP spoof attack, and as a result the anomaly was detected by the corresponding ICMP and DNS flow workers. 


These observations can help to determine a weight for individual workers when identifying attack flow(s). For example, if ARP and local TCP port 80 workers flag an anomaly simultaneously, then it is worth investigating bidirectional TCP flows to/from port 80 -- deprioritizing alarms from the ARP worker. 


\subsection{Isolating Attack Microflows}\label{sec:attack5tuple}
We now analyze the performance of identifying attack microflows (a group of microflows) and the associated cost when stage-3 inferencing is introduced. Table~\ref{table:stage3performance} shows the result for two scenarios: (i) full structure of inferencing is operational, \ie stage-3 is dynamically activated by alarms of stage-1 and stage-2, (ii) only stage-3 is permanently operational.  

\textbf{(i) Dynamic Activation of Stage-3:} Once stage-3 inferencing is dynamically added to the pair of stage-1 and stage-2, the FPR drops to $1.2$\%  for the aggregate of all IoT devices-- this FPR is a third of the best result achieved without stage-3 models, reported in Table~\ref{table:performance}. Focusing on individual device types, the highest FPR for the Chromecast improves from $19.7$\% to $5$\% -- stage-3 models correct more than 10\% of false alarms generated by the pair of stage-1 and stag-2 -- a similar improvement in FPR is also observed for all other device types. Overall, use of the full structure (pair of stage-1 and stage-2 with dynamic activation of stage-3) of inference models has improved FPR by $4$\% with a minor drop in TPR  compared to the case when only the pair of stage-1 and stag-2 is employed, but our scheme was able to successfully detect both direct and distributed attacks at some stage of their lifetime. This shows the collective power of knowledge learned by individual models at the three stages of inferencing. Of course, a combination of the three inferencing engines gives the best performance, but it comes at cost of additional computing which needs to be managed judiciously. Therefore, stage-3 gets activated only when both stage-1 and stage-2 raise anomaly flags, and hence the expensive compute of inferencing at stage-3 is needed on-demand and only for a subset of the network traffic corresponding to specific device(s) and their anomalous service flows
-- combating distributed attacks that target multiple devices simultaneously would require a slightly different strategy which is beyond the scope of this paper. Use of the stage-3 inferencing enabled us to identify contributing microflows in all direct attacks as well as highest-matching flows in all distributed attacks that we launched.

\begin{table}
	\centering
	\caption{Cost of inspecting regular packets for inferencing at stage-3.}
	\label{table:packetInspectioncost-devs}
	\vspace{-3mm}
	\begin{adjustbox}{max width=0.48\textwidth}	
		\renewcommand{\arraystretch}{1.2}			
		\begin{tabular}{|l|c|}
			\hline 
			IoT device & inspected packet \%\tabularnewline
			\hline 
			WM & 9.3\tabularnewline
			\hline 
			TP & 14.3\tabularnewline
			\hline 
			SC & 9.9\tabularnewline
			\hline 
			NC & 10.1\tabularnewline
			\hline 
			CU & 30.4\tabularnewline
			\hline 
			AE & 45.4\tabularnewline
			\hline 
			All devices & 26.0\tabularnewline
			\hline 
		\end{tabular}
	\end{adjustbox}	
	\vspace{-6mm}
\end{table}

\textbf{(ii) Permanent Use of Stage-3:} When only stage-3 inferencing is permanently used, the TPR across the aggregate of all devices is $91.3$\% which is slightly better than the TPR $89.7$\% achieved by the pair stage-1 and stage-2, shown in Table~\ref{table:performance}.
Moving to individual devices, the largest improvement of TPR is observed in Netatmo Camera (NC), rising from $80.3$\% to $95.9$\%. In addition to better detection, stage-3 inferencing provides a finer grain visibility into the attack traffic. Considering false alarms, however, the stage-3 yields an FPR of $8.6$\% (for aggregate of all devices) which is almost double that of the pair of stage-1 and stage-2. Giving a high FPR is seen across all individual IoT devices, peaking at $32.3$\% in WeMo motion (WM). For this main reason we leverage the combination of all three stages in order to achieve a minimum FPR while having an acceptable TPR.

\begin{figure}[t!]
	\centering
	\includegraphics[width=0.45\textwidth, height=0.27\textwidth]{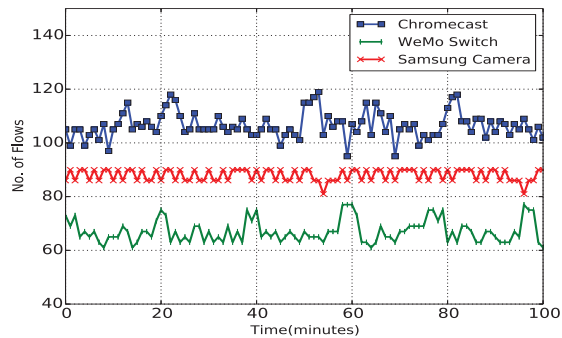}
	\vspace{-3mm}
	\caption{Number of entries in the SDN switch flow-table when only stage-3 inferencing is permanently used (for regular benign traffic).}
	\label{fig:tuplemonitor}
	\vspace{-6mm}
\end{figure}

In terms of processing cost, we note that when stage-3 is permanently operational, packet inspection is performed even for regular packets (which may not necessarily be malicious) to identify microflows. 
Also, insertion of microflows into the SDN switch (to suppress mirroring packets) will consume TCAM table entries. Table~\ref{table:packetInspectioncost-devs} shows the fraction of packets (benign traffic only) that undergo headers inspection. We see that $26.0$\% of packets (in our set of IoTs) across the aggregate of all devices get inspected. Obviously, this fraction variety from a device to another depending on the dynamics of their network activity -- Wemo motion (WM) has the lowest fraction of $9.3$\% and Amazon Echo (AE) has the highest fraction of $45.4$\%. On the other hand, when the pair of stage-1 and stag-2 is in charge of inferencing, only less then $1$\% of packets are inspected (purely to extract DNS bindings). Moving to TCAM entries, we show in Fig.~\ref{fig:tuplemonitor} the real-time (minutely) count of flow entries needed for three representative devices (Chromecast, WeMo switch, and Samsung camera).
Note that the count fluctuates over time since microflow entries are configured with an idle-timeout, while MUD service flows stay permanently in the TCAM table. It can be seen that on average 91, 66, and 88 entries are consumed, respectively to monitor the network activity of Chromecast, WeMo switch, and Samsung camera when stage-3 is permanently used. These numbers are almost double those (\ie 36, 30, 45) of when only the pair of stage-1 and stage-2 is used. From cost reduction perspective, it would be beneficial  for our system to make use of stage-3 dynamically and on-demand in conjunction with the pair of stage-1 and stage-2.

\setlength{\figurewidthA}{0.45\textwidth}
\setlength{\figureheigthA}{0.2\textwidth}
\begin{figure*}[t!]
	\centering
	\subfigure[varying source IP address.] 
	{
		\includegraphics[width=\figurewidthA,height=\figureheigthA]{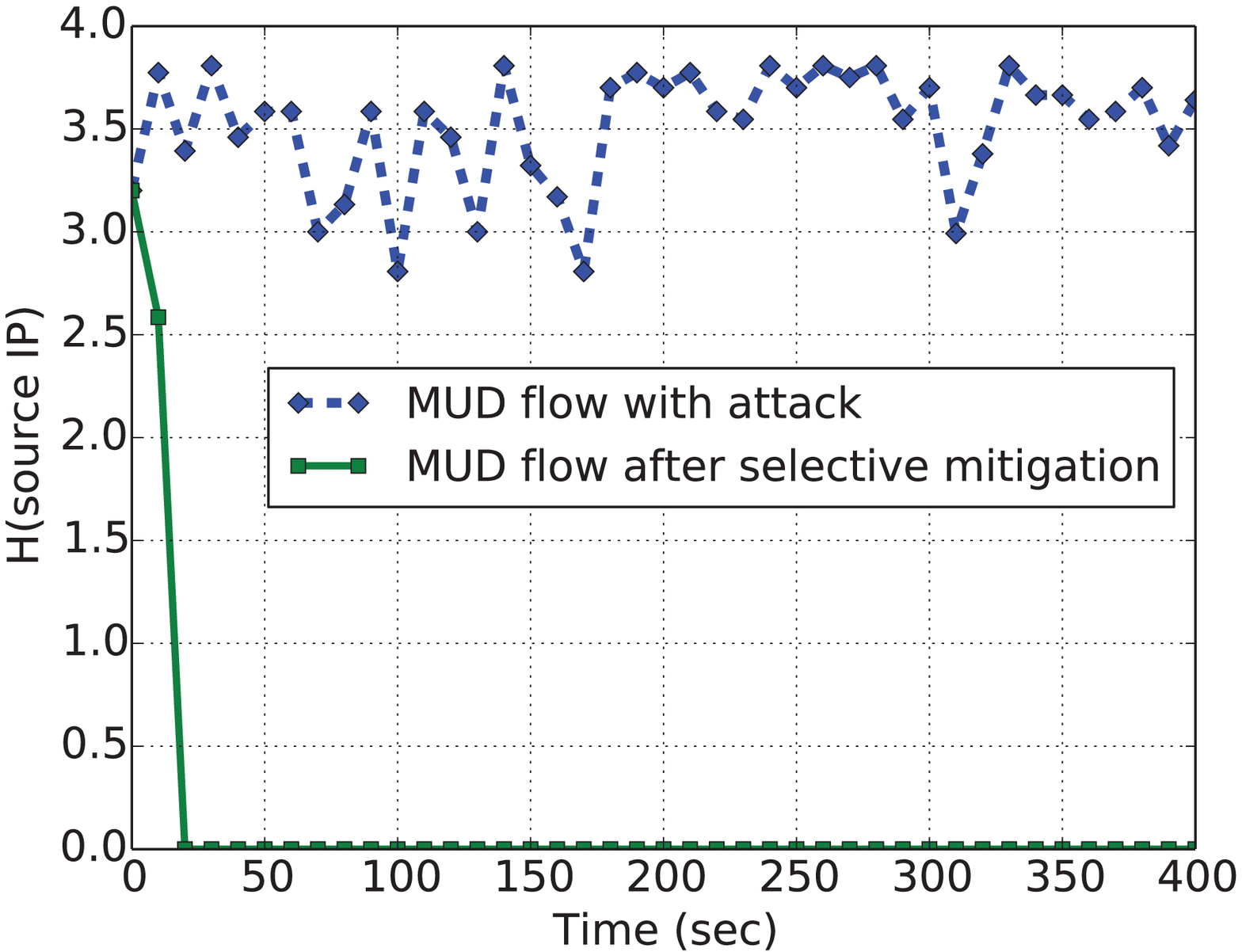}
		\label{fig:ipAttack}
	}
	\subfigure[varying source port number.] 
	{
		\includegraphics[width=\figurewidthA,height=\figureheigthA]{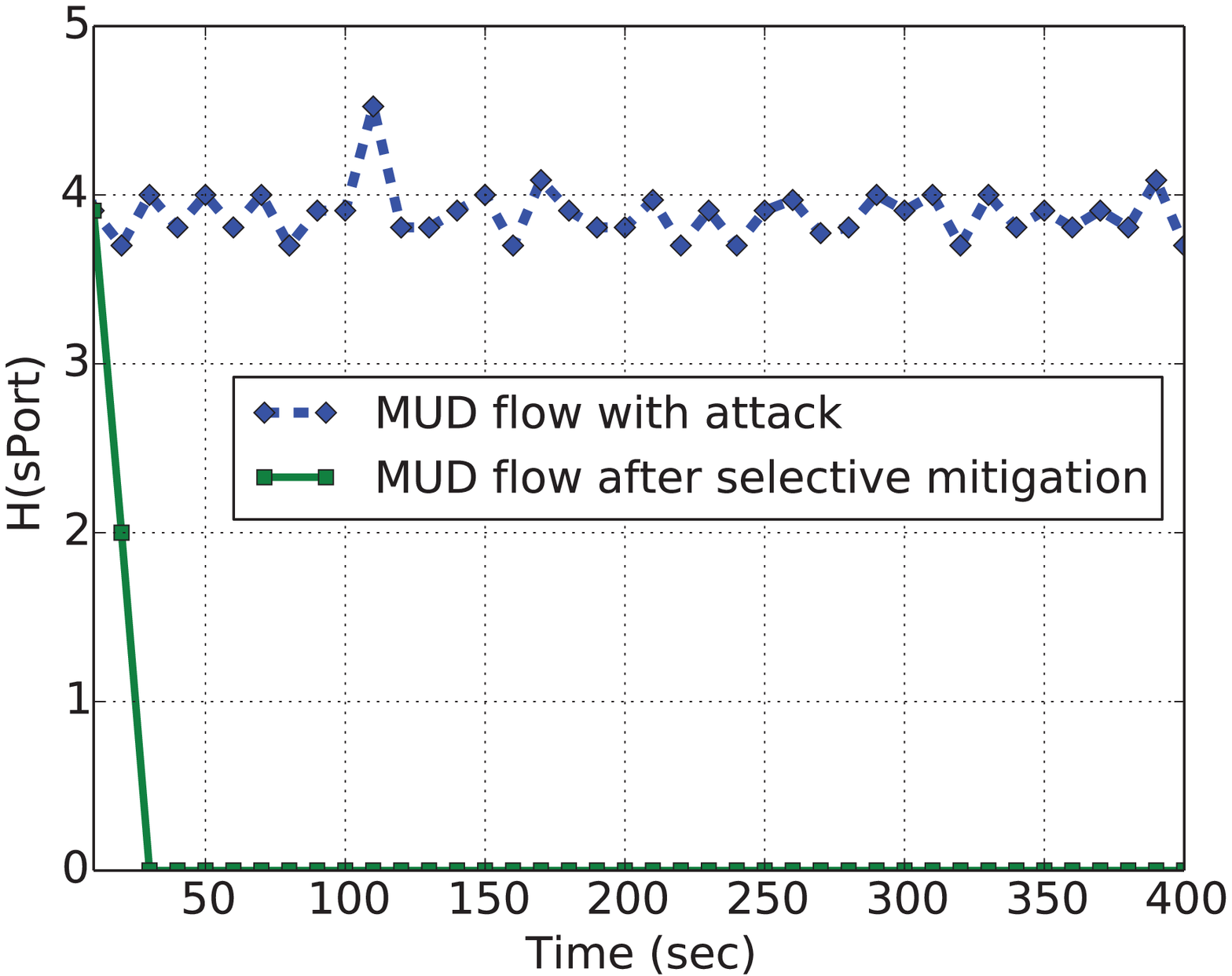}
		\label{fig:portAttack}
	}
	\vspace{-2mm}
	\caption{Time-trace of entropy of microflow headers in distributed attacks: (a) varying source IP, and (b) varying source port number, on WeMo switch over MUD flow TCP 49153.}
	\label{fig:entropychangeduringattack}
	\vspace{-6mm}
\end{figure*}

\textbf{Dispersion Anomaly Detection and Mitigation:}
We now look at the performance of our inferencing as well as mitigation strategy in distributed attacks. Fig.~\ref{fig:entropychangeduringattack} illustrates the time-trace of entropy of microflow headers in two distributed attacks (dashed blue lines) launched on WeMo switch over its MUD flow TCP 49153 -- each displays a high entropy in one header (sIP or sPort) of microflows. In both cases, it can be seen that once a distributed attack is detected by the dispersion anomaly worker, and thereby mitigated, the entropy slightly drops (solid green lines, from $3.2$ to $2.6$ in Fig.~\ref{fig:ipAttack} and from $3.9$ to $2.0$ in Fig.~\ref{fig:portAttack}), but it is far above zero -- meaning that the slow-changing portion of the distributed attack is still present. However, this portion of attack is completely mitigated in the second attempt.

\begin{table*}[!t]
	\begin{minipage}{.68\linewidth}
		\caption{Comparing our solution with Snort in detecting attacks.}
		\centering
		\label{table:snortresults}
		\vspace{-0.3cm}
		\begin{adjustbox}{max width=0.8\textwidth}	
			\renewcommand{\arraystretch}{1.2}			
			\begin{tabular}{|l|>{\raggedright}p{7cm}|>{\raggedright}p{6.3cm}|}
				\hline 
				\multirow{1}{*}{\textbf{IoT device}} & \multirow{1}{*}{\textbf{Detected wild attackers (by our solution)}} & \textbf{Detected attacks (by Snort)}\tabularnewline
				\hline 
				WM & \{107.170.227.13\} & \{107.170.227.13\}, {\color{blue}SSDP reflection(I$\rightarrow$d$\rightarrow$I)}\tabularnewline
				\hline 
				WS & \{107.170.228.161\} & \{107.170.228.161\}\tabularnewline
				\hline 
				SC & {\{103.29.71.94, 45.55.2.34, 107.170.229.67, 45.55.14.102, {\color{red}181.214.206.55, 216.98.153.254, 54.215.173.102, 14.134.5.4, 205.209.159.120}\}} & \{103.29.71.94, 45.55.2.34, 107.170.229.67,
				45.55.14.102\}, {\color{blue}SNMP reflection(I$\rightarrow$d$\rightarrow$I)}\tabularnewline
				\hline 
				TP & \{107.170.226.164,
				185.170.42.66, 46.182.25.42, 45.227.254.243, {\color{red}185.156.177.13, 17.136.0.172, 125.212.217.214, 107.170.225.175, 217.182.197.186}\} & \{107.170.226.164, 185.170.42.66, 46.182.25.42, 45.227.254.243\}\tabularnewline
				\hline 
				NC & \{58.182.245.89, 27.75.133.76, 14.234.90.16, 103.4.117.85,   {\color{red}177.74.184.229, 176.36.241.230, 81.17.18.221, 201.174.9.186, 194.208.107.25, 161.97.195.135,
					189.165.40.237}\} & \{58.182.245.89, 27.75.133.76, 14.234.90.16, 103.4.117.85\}\tabularnewline
				\hline
				CU, PH & N/A  & {\color{blue}SSDP reflection(I$\rightarrow$d$\rightarrow$I)} \tabularnewline
				\hline
				AE, IH, LX & N/A &  \tabularnewline
				\hline 
			\end{tabular}
		\end{adjustbox}
	\end{minipage}%
	\begin{minipage}{.32\linewidth}
		\centering
		
		\caption{Performance of other anomaly detectors.\label{table:machinelearniners}}
		\vspace{-0.3cm}
		\begin{adjustbox}{max width=0.80\textwidth}	
			\renewcommand{\arraystretch}{1.2}			
			\begin{tabular}{|l|c|c|}
				\hline 
				\multirow{1}{*}{\textbf{IoT device}} & \textbf{TPR (\%)} & \textbf{FPR (\%)}\tabularnewline
				\hline 
				WeMo motion & 74.00 & 3.32\tabularnewline
				\hline 
				WeMo  switch& 60.55 & 1.10\tabularnewline
				\hline 
				Samsung smartcam & 67.78 & 2.31\tabularnewline
				\hline 
				TP-Link smart plug & 95.19 & 0.63\tabularnewline
				\hline 
				Netatmo camera & 0.00 & 0.10\tabularnewline
				\hline
				Chromecast Ultra & 32.93 & 3.35\tabularnewline
				\hline 
				Amazon Echo& 15.18 & 3.27\tabularnewline
				\hline 
				Phillips Hue bulb & 19.86 & 3.26\tabularnewline
				\hline 
				iHome Smart plug & 70.00 & 1.94\tabularnewline
				\hline 
				LiFX bulb & 82.00 & 2.52\tabularnewline
				\hline 
			\end{tabular}
		\end{adjustbox}
	\end{minipage} 
	\vspace{-3mm}
\end{table*}

\vspace{-3mm}
\subsection{Comparison with Existing Methods}\label{sec:comparison}
Lastly, we compare the performance of our scheme with existing tools and proposals. We start with Snort \cite{snort}, a widely deployed, open-source, signature based IDS, and then reevaluate our models with the features proposed by other researchers.

\textbf{Snort IDS:} We configured Snort IDS with the community rule-set \cite{snortrules} and replayed our packet traces to Snort IDS using the {\myverb{tcpreplay}} tool. Table~\ref{table:snortresults} lists the IP address of endpoints on the Internet that attacked our testbed during the experiments, and were detected by our specification-based intrusion detector because these Internet endpoints were not specified by the MUD profile of the IoT devices. These wild attacks from the Internet were seen after port forwarding was enabled on the gateway. According to AbuseIPDB\cite{abuseipdb}, most of these endpoints have been reported as abusive IP addresses (\eg 181.214.206.55 has a probability of 46\% as being an abusive IP address). We can see that the Snort detects a subset of these attacks -- attacks from IP addresses in red text are not detected by the Snort. In addition, out of 40 types of intended attacks, the Snort detected only two, namely SSDP reflection (I$\rightarrow$d$\rightarrow$I) on WeMo motion and SNMP reflection (I$\rightarrow$d$\rightarrow$I) on Samsung camera, shown by blue text in Table~\ref{table:snortresults}. These two types of attacks (detected by Snort) carry traffic towards the local network and their signature was known to Snort, whereas the majority of the intended attacks were specifically designed for IoT devices, and Snort does not have the signature for most of them.

\textbf{Other machine learners:} Works in \cite{braga2010lightweight,cui2016sd,tang2016deep,bhunia2017dynamic} also use a machine learning based approach to detect anomalies. However, the main issue of their approach is that their models are based on binary classification and use both benign and attack traffic for the training, which limits the scalability of using such methods in an operational network. We note they also employ packet/byte counters as the feature for training their models, but at device-level only (\ie two features: aggregate bytes and packets of all flows). 
To demonstrate the superiority of our approach, the anomaly detection process was modified to use only these two attributes, and the results are illustrated in Table~\ref{table:machinelearniners}.
It is apparent that the overall performance (across 5 devices) is very poor compared to our scheme, with no attacks detected for the Netatmo as well as half of the attacks are missed for the WeMo motion and Samsung smartcam devices. In addition, this single model approach does not provide any indication of attack flows.

\section{Management of Machine Learning Models}\label{sec:ml} 

\begin{table*}
	\centering
	\caption{Dataset summary and accuracy on managing machine learning models. }
	\label{table:datasetml}
	\vspace{-3mm}
	\begin{adjustbox}{max width=0.98\textwidth}
		\renewcommand{\arraystretch}{1.2}				
		\begin{tabular}{|l|c|c|c|c|c|c|c|c|c|c|}
			\hline 
			\multirow{2}{*}{Device types} & \multirow{2}{*}{\# devices} & \multirow{2}{*}{total \# data instances} & \multicolumn{2}{c|}{model per IoT unit} & \multicolumn{2}{c|}{model per IoT type [naive]} & \multicolumn{2}{c|}{model per IoT type [universal]} & \multicolumn{2}{c|}{model per IoT type [progressive]}\tabularnewline
			\cline{4-11} 
			&  &  & \# train instances & Accuracy & \# train instances & Accuracy (min-max) & \# train instances & Accuracy & \# train instances  & Accuracy\tabularnewline
			\hline 
			Cisco cam & 45 & 6,248,812 & 675,000 & \textbf{93.0} & 15000 & 31.8 - 53.2 & 675,000 & \textbf{98.9} & 44,591 & 98.9\tabularnewline
			\hline 
			Axis cam & 8 & 1,109,648 & 120,000 & 89.4 & 15,000 & 19.6 - 51.5 & 120,000 & 92.7 & 47,853 & 92.0\tabularnewline
			\hline 
			Steinel cam & 6 & 276,892 & 90,000 & 91.6 & 15,000 & 65.7 - 96.4& 90,000 & 96.1 & 20,949 & \textbf{93.8}\tabularnewline
			\hline 
			WeMo motion & 4 & 160,427 & 60,000 & 91.6 & 15,000 & 21.2 - 88.4 & 60,000 & 96.2 & 24,312 & 95.3\tabularnewline
			\hline 
			Samsung cam & 4 & 213,881 & 60,000 & 97.5 & 15,000 & 41.2 - 64 & 60,000 & 98.3 & 30,865 & 96.3\tabularnewline
			\hline 
			Chromecast & 4 & 143,446 & 60,000 & \textbf{84.2 }& 15,000 & 49.8 - 73.2& 60,000 & \textbf{92.1} & 32,685 & 90.9\tabularnewline
			\hline 
			Alexa & 4 & 191,036 & 60,000 & 93.2 & 15,000 & 57.7 - 75.1 & 60,000 & 94.5 & 26,175 & 94.5\tabularnewline
			\hline 
			Netatmo cam & 4 & 189,628 & 60,000 & 96.9 & 15,000 & \textbf{56.2 - 96.4}& 60,000 & 98.7 & 24,423 & 98.7\tabularnewline
			\hline 
			TPlink plug & 4 & 213,186 & 60,000 & 98.7 & 15,000 & \textbf{95.48 - 98.2} & 60,000 & 99.1 & 16,846 & 99.1\tabularnewline
			\hline 
			
		\end{tabular}
		
	\end{adjustbox}	
	\vspace{-4mm}
\end{table*}

\begin{figure}[t!]
	\centering
	\includegraphics[width=0.45\textwidth,height=0.27\textwidth]{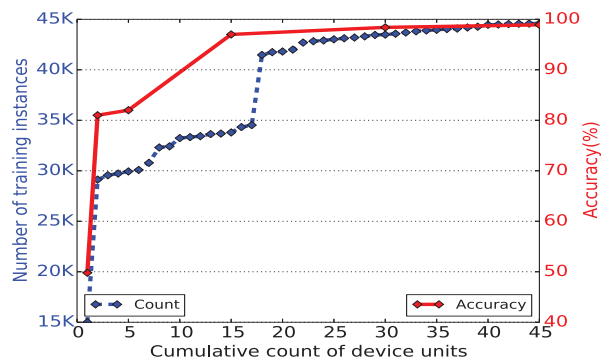}
	\vspace{-3mm}
	\caption{Accuracy and training dataset for model generated using different counts of Cisco Cameras.}
	\label{fig:ciscoCameraCount}
	\vspace{-4mm}
\end{figure}

\begin{table*}
	\centering
	\caption{Summary on model size and training time.}
	\label{table:datasetmlcost}
	\vspace{-3mm}
	\begin{adjustbox}{max width=0.7\textwidth}
		\renewcommand{\arraystretch}{1.2}				
		\begin{tabular}{|l|c|c|c|c|c|c|}
			\hline 
			\multirow{2}{*}{Device types} & \multicolumn{2}{c|}{model per IoT unit} & \multicolumn{2}{c|}{model per IoT type [universal]} & \multicolumn{2}{c|}{model per IoT type [progressive]}\tabularnewline
			\cline{2-7} 
			& model size (MB) & training time (s) & model size (MB) & training time (s) & model size (MB) & training time(s)\tabularnewline
			\hline 
			Cisco cam & 41 & \textbf{6233} & 1 & \textbf{20444} & 1 & \textbf{317}\tabularnewline
			\hline 
			Axis cam & 8 & 1222 & 2 & 1440 & 2 & 313\tabularnewline
			\hline 
			Steinel cam & 1 & 44 & 1 & 21 & 1 & 17\tabularnewline
			\hline 
			WeMo motion & 15 & 427 & 4 & 1073 & 4 & 373\tabularnewline
			\hline 
			Samsung cam & 34 & 339 & 9 & 467 & 9 & 187\tabularnewline
			\hline 
			Chromecast & 13 & 1095 & 4 & 2182 & 4 & 743\tabularnewline
			\hline 
			Alexa & 6 & 348 & 2 & 482 & 2 & 310\tabularnewline
			\hline 
			Netatmo cam & 6 & 303 & 2 & 401 & 2 & 142\tabularnewline
			\hline 
			TPlink plug & 4 & 41 & 1 & 54 & 1 & 10\tabularnewline
			\hline 
		\end{tabular}
	\end{adjustbox}	
	\vspace{-4mm}
	
\end{table*}	

In this section, we discuss the scalability of our inferencing scheme in networks with a large number of connected devices  by considering various strategies and challenges to manage machine learning models.

\textbf{Dataset:} We collected flow records of various IoT device types in two different environments: (a) four home networks consisting of consumer IoTs, and (b) a subset of an enterprise IoT network.
For the home network scenario, we replicated our SDN-based lab setup, shown in Fig.~\ref{fig:testbed}, in three homes (of this paper's authors) for a duration of a week. Each home setup consisted of an SDN-enable home gateway and six IoT devices including WeMo motion, Samsung camera, Chromecast, Alexa, Netatmo camera, and TPlink plug. The home gateway was controlled by the SDN controller located in our university lab. 
For the enterprise scenario, the IT department of our university provisioned a full mirror (both inbound and outbound) of traffic corresponding to a subset of the campus IP camera network to our data collection system. We obtained ethics clearance (UNSW Human Research Ethics Advisory Panel approval number HC190171) for this experiment -- our data collection system only records flow-level counters traffic for connected cameras of three types Cisco camera, Axis camera, and Steinel camera.
The first three columns in Table~\ref{table:datasetml} summarize our dataset by indicating IoT device types, number of units per each device type, and total number of flow-level instances (volumteric features computed over a minute) collected in our dataset. 


\textbf{Strategies for Generating Models:} We can take various strategies to generate and maintain models of IoT devices for a large-scale deployment of our scheme. The first strategy is to generate a specific model per each unit of IoT devices. This approach causes our models to grow linearly by the count of devices and size of their instances. The column ``model per IoT unit'' in Table~\ref{table:datasetml} shows the total number of training instances per each IoT type and the average accuracy achieved  across individual unit models of that type in correctly classifying instances (benign and attack -- except in Cisco camera and Axis camera for which we had only benign data). The second strategy is ``federated learning'' where we maintain one model per each type of IoT devices. This strategy can be implemented in three ways: (i) the model of a randomly chosen unit from a given device type is used for all units of that type. This approach is captured by column ``model per IoT type [naive]'' in Table~\ref{table:datasetml} for which 15000 instances were used to generate each unit model, representing the model of a device type. This way the linear growth of models is managed, but the accuracy, \ie min-max value pair, is highly variable (almost unacceptable) depending on which unit model is chosen to represent a given IoT type; (ii) a universal model is generated and maintained for a device type by mixing instances from individual units of that type. The column ``model per IoT type [universal]'' in Table~\ref{table:datasetml} shows the prediction accuracy which is far better than that of the naive approach across various models; and (iii) the universal approach is slightly modified where the model is progressively updated by iterating over individual units of a given type. A universal model is first built by instances of a unit, and tested against instances of another unit of that type. The model gets re-trained by additional instances of the tested unit that are misclassified by the model. This process repeats till all units get covered, and hence the universal model progressively learns unique behaviors of a device type across units. The accuracy results under column ``model per IoT type [progressive]'' in Table~\ref{table:datasetml} are almost comparable to the universal approach. Instead, the number training instances has reduced significantly -- for example, in Cisco camera model, it has dropped from 675000 instances (1.85 GB) to 44,591 instances (123 MB). Fig~\ref{fig:ciscoCameraCount} shows the progression of prediction accuracy and training instances count as functions of cumulative count of units covered in progressively developing the universal model of Cisco camera. It can be seen that the accuracy gets almost saturated once 15 units of camera are covered. Obviously, the evolution may slightly change if we change the order of units considered in the process, but the trend of both curves would be always non-decreasing.

\textbf{Model size and training time:} Lastly, Table~\ref{table:datasetmlcost} shows the model size and the training time for three strategies that give accepted levels of accuracy (as per Table~\ref{table:datasetml}). The unit-based strategy would obviously require more memory to store models, compared to type-based (federated) strategies. For example, $41$ MB of storage is needed for 45 unit-based models of Cisco camera while the size of a universal model is $1$ MB. 
In terms of training time, we note that time to train a model increases by the number of its training instances. Therefore, generating a universal model would take longer time compared to a unit-based model. Focusing on Cisco camera as a representative type, to generate 45 unit-based models (not in parallel) it takes about 2 hours, and the basic universal model alone needs about 5.5 hours to be built while this training time is significantly reduced to 5 minutes when a progressive strategy is chosen to generate the universal model. 

\section{Conclusions}\label{sec:conclusion} 
Vulnerable IoT devices are increasingly putting smart environments at risk by exposing their networks unprotected to cyber attackers. MUD framework aims to reduce the attack surface on IoTs by formally defining their expected network behavior. In this paper, we have focused on detecting IoT microflows involved in various volumteric and distributed attacks that are not prevented by MUD profile enforcement. We developed an SDN-based system empowered by machine learning to monitor and learn the behavioral patterns of MUD rules at multiple levels of granularity. We then subjected real IoT devices to a range of volumetric attacks (designed to conform to MUD profiles) in our lab, collected and labeled our traffic traces. we prototyped our system and quantified the efficacy of our scheme in detecting volumetric attacks. We made our dataset and system available to the public. Lastly, we demonstrated various strategies in managing machine learning models in large scale deployments of our scheme, and analyzed their accuracy versus training cost.

\bibliographystyle{IEEEtran}
\bibliography{mudMitigate}

\vspace{-1cm}
\begin{IEEEbiography}[{\includegraphics[width=1in,height=1.25in,clip,keepaspectratio]{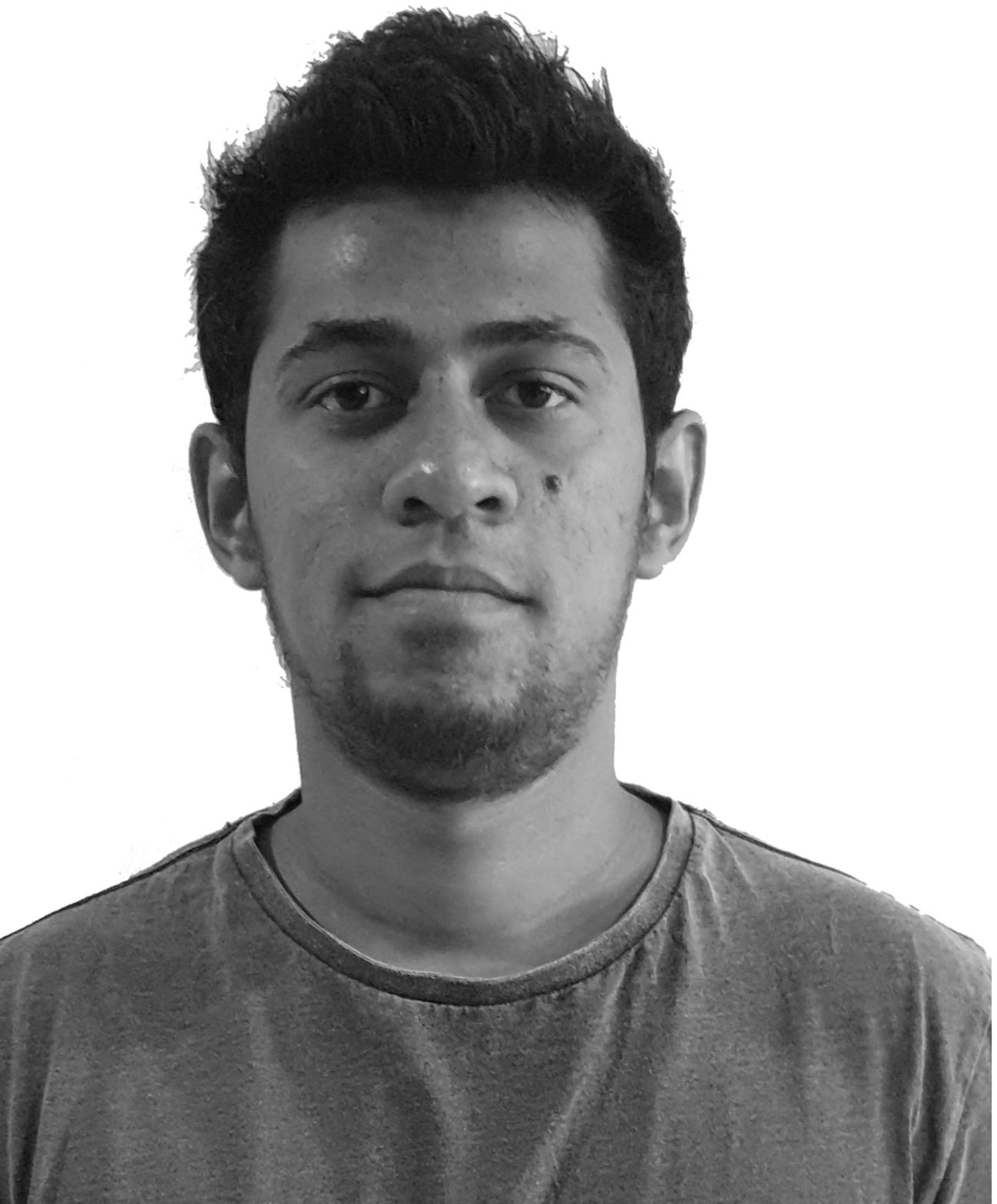}}]{Ayyoob~Hamza}
	received his Bachelors' degree in Computer Science from the University of Colombo, Sri Lanka and is currently a Ph.D. Candidate at the University of New South Wales in Sydney, Australia. Prior to his research career, he worked at WSO2 Inc. as a Senior Software Engineer for 3 years working on IoT solutions. His research interests includes Internet of Things, Network Security, Distributed Systems and Software-Defined Networking.
\end{IEEEbiography}

\vspace{-1cm}
\begin{IEEEbiography}[{\includegraphics[width=1in,height=1.25in,clip,keepaspectratio]{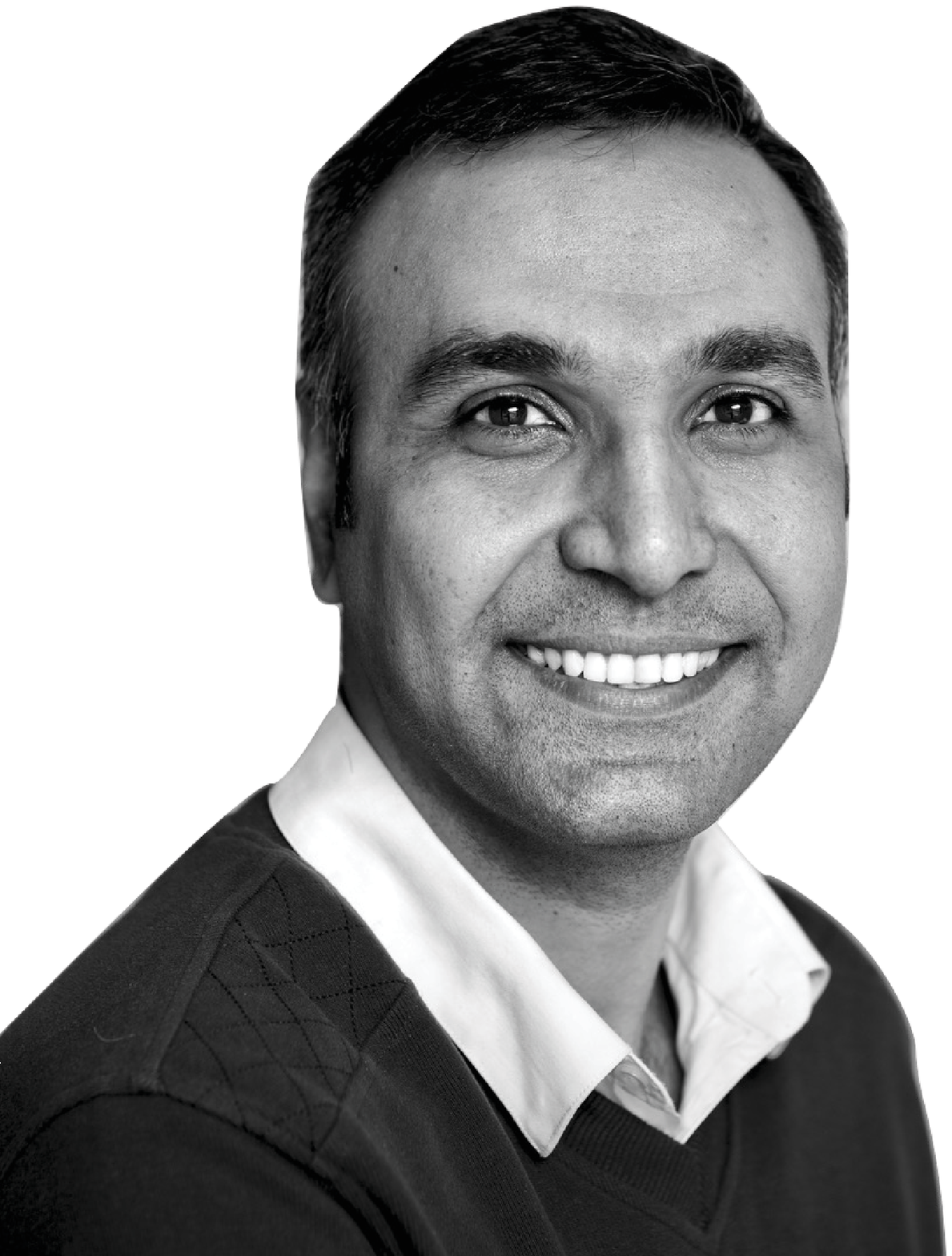}}]{Hassan~Habibi~Gharakheili}
	received his B.Sc. and M.Sc. degrees of Electrical Engineering from the Sharif University of Technology in Tehran, Iran in 2001 and 2004 respectively, and his Ph.D. in Electrical Engineering and Telecommunications from UNSW in Sydney, Australia in 2015. He is currently a Senior Lecturer at UNSW Sydney. His current research interests include programmable networks, learning-based networked systems, and data analytics in computer systems.
\end{IEEEbiography}

\vspace{-1cm}
\begin{IEEEbiography}[{\includegraphics[width=1in,height=1.25in,clip,keepaspectratio]{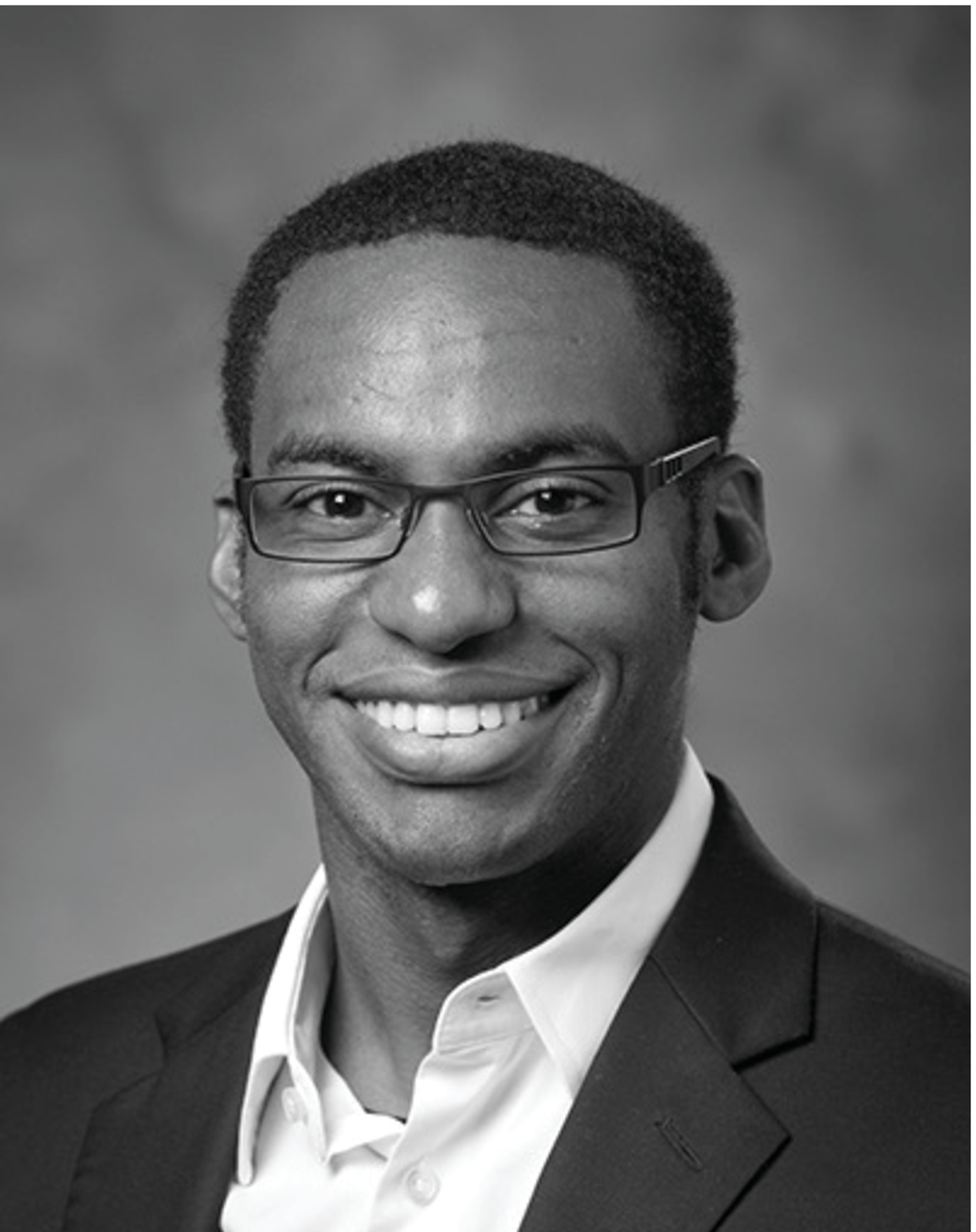}}]{Theophilus~A.~Benson}
	Theophilus Benson received his Ph.D. from University of Wisconsin, Madison in 2012 and his B.S. from Tufts university in 2004.  He is now a Professor at Carnegie Mellon University. His research focuses on designing frameworks and algorithms for solving practical networking problems with an emphasis on speeding up the internet, improving network reliability,  and simplifying network management.  
\end{IEEEbiography}

\begin{IEEEbiography}[{\includegraphics[width=1in,height=1.25in,clip,keepaspectratio]{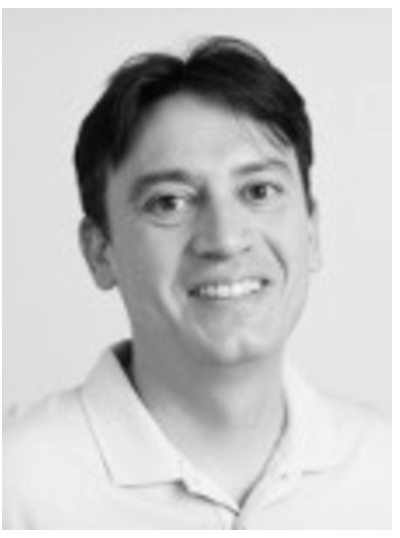}}]{Gustavo~Batista}
	received his MSc and PhD degrees in Computer Science from the University of Sao Paulo, Brazil, in 1997 and 2003, respectively. He was an associate professor at the University of Sao Paulo (2007-2018) and an associate professor at the University of New South Wales (UNSW). His research interests fall in Machine Learning, including time series and data stream quantification and classification.
\end{IEEEbiography}

\vspace{-1cm}
\begin{IEEEbiography}[{\includegraphics[width=1in,height=1.25in,clip,keepaspectratio]{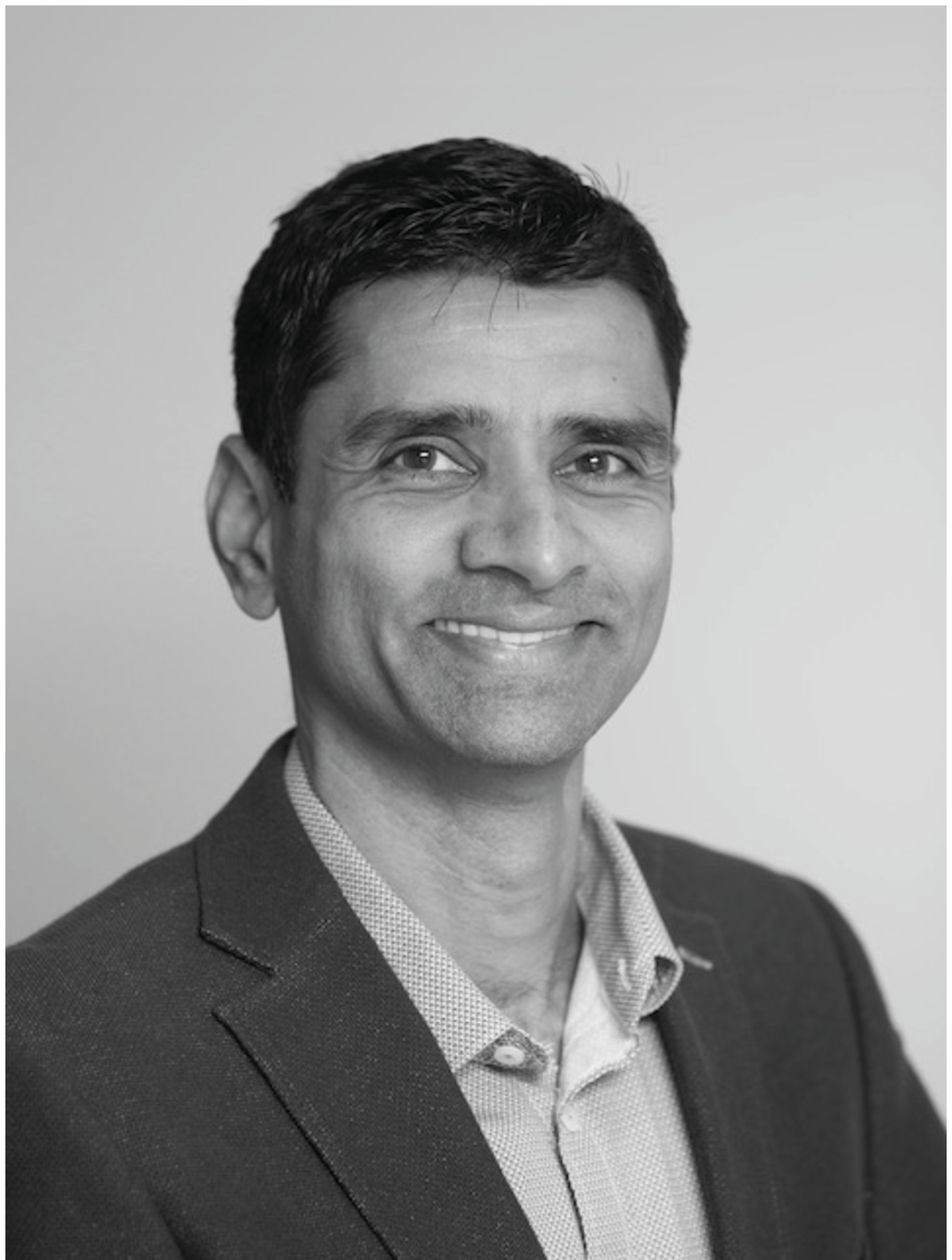}}]{Vijay Sivaraman}
	received his B. Tech. from the Indian Institute of Technology in Delhi, India, in 1994, his M.S. from North Carolina State University in 1996, and his Ph.D. from the University of California at Los Angeles in 2000. He has worked at Bell-Labs as a student Fellow, in a silicon valley start-up manufacturing optical switch-routers, and as a Senior Research Engineer at the CSIRO in Australia. He is now a Professor at the University of New South Wales in Sydney, Australia. His research interests include Software Defined Networking, network architectures, and cyber-security particularly for IoT networks.
\end{IEEEbiography}

\balance

\end{document}